\documentclass[conference]{IEEEtran}
\IEEEoverridecommandlockouts
\DeclareRobustCommand{\IEEEauthorrefmark}[1]{\smash{\textsuperscript{\footnotesize #1}}}
\usepackage{cite}

\usepackage{amsmath,amssymb,amsfonts}
\usepackage{algorithmic}
\usepackage[hidelinks]{hyperref}
\usepackage{url}
\usepackage{flushend}

\newcommand{\conditionalURL}[1]{%
  \ifdefined\DOUBLEBLIND
    URL withheld to preserve blind review.%
  \else
    \url{#1}%
  \fi
}
\newcommand{\conditionalCite}[1]{%
  \ifdefined\DOUBLEBLIND
    \cite{#1}%
  \else
    \cite{#1}%
  \fi
}

\usepackage{graphicx}
\usepackage{textcomp}
\usepackage{comment}
\usepackage{xcolor}
\usepackage{subfig} % for the two SSD figures
\usepackage{graphics} % for hspace in SSD figure placement
\usepackage{anyfontsize} % for font size change of table
\usepackage[font=small,labelfont=small]{caption}
\usepackage[capitalise]{cleveref}
\def\BibTeX{{\rm B\kern-.05em{\sc i\kern-.025em b}\kern-.08em
    T\kern-.1667em\lower.7ex\hbox{E}\kern-.125emX}}
\usepackage{orcidlink} %ORCID logo with link

\begin{document}

\title{PowerSensor3: A Fast and Accurate Open Source Power Measurement Tool
\ifdefined\DOUBLEBLIND
\else
\thanks{
This work received funding from the European Union through the RADIOBLOCKS (101093934) and MCSA-RISE Cloudstars (101086248) projects, the Dutch Research Council (NWO) through the DAS-6 (621.018.201), CORTEX (NWA.1160.18.316), OffSense (OCENW.KLEIN.209) and MLS (OCENW.KLEIN.561) grants, and from the Netherlands eScience Center through the RECRUIT (ETEC.2020.025) grant. K. Doekemeijer is funded by the VU PhD innovation program.}
\fi
}

\ifdefined\DOUBLEBLIND
\author{author list removed for double blind review}
\else
\author{
    \IEEEauthorblockN{Steven van der Vlugt\IEEEauthorrefmark{1}\orcidlink{0000-0001-6834-4860}, Leon Oostrum\IEEEauthorrefmark{2}\orcidlink{0000-0001-8724-8372}, Gijs Schoonderbeek\IEEEauthorrefmark{1}\orcidlink{0000-0001-9482-1253}, 
Ben van Werkhoven\IEEEauthorrefmark{3,2}\orcidlink{0000-0002-7508-3272 }, \\ Bram Veenboer\IEEEauthorrefmark{1}\orcidlink{0000-0001-9607-1142},
    Krijn Doekemeijer\IEEEauthorrefmark{4}\orcidlink{0009-0007-7530-4438}, John W. Romein\IEEEauthorrefmark{1}\orcidlink{0000-0002-1915-5067}}
    \IEEEauthorblockA{\IEEEauthorrefmark{1}ASTRON (Netherlands Institute for Radio Astronomy), Dwingeloo, the Netherlands
    \\\{vlugt, schoonderbeek, veenboer, romein\}@astron.nl}
    \IEEEauthorblockA{\IEEEauthorrefmark{2}Netherlands eScience Center, Amsterdam, the Netherlands, l.oostrum@esciencecenter.nl}
    \IEEEauthorblockA{\IEEEauthorrefmark{3}Leiden University, Leiden, the Netherlands, b.van.werkhoven@liacs.leidenuniv.nl}
    \IEEEauthorblockA{\IEEEauthorrefmark{4}Vrije Universiteit Amsterdam, Amsterdam, the Netherlands,
    k.doekemeijer@vu.nl}
}
\fi

\maketitle

% force page numbers
\thispagestyle{plain}
\pagestyle{plain}

\begin{abstract}
Power consumption is a major concern in data centers and HPC applications, with GPUs typically accounting for more than half of system power usage. 
While accurate power measurement tools are crucial for optimizing the energy efficiency of (GPU) applications, both built-in power sensors as well as state-of-the-art power meters often lack the accuracy and temporal granularity needed, or are impractical to use.
Released as open hardware, firmware, and software, PowerSensor3 provides a cost-effective solution for evaluating energy efficiency, enabling advancements in sustainable computing. 
The toolkit consists of a baseboard with a variety of sensor modules accompanied by host libraries with C++ and Python bindings.
PowerSensor3 enables real-time power measurements of SoC boards and PCIe cards, including GPUs, FPGAs, NICs, SSDs, and domain-specific AI and ML accelerators.
Additionally, it provides significant improvements over previous tools, such as a robust and modular design, current sensors resistant to external interference, simplified calibration, and a sampling rate up to 20\,kHz, which is essential to identify GPU behavior at high temporal granularity.
This work describes the toolkit design, evaluates its performance characteristics, and shows several use cases (GPUs, NVIDIA Jetson AGX Orin, and SSD), demonstrating PowerSensor3's potential to significantly enhance energy efficiency in modern computing environments.

\end{abstract}


\section{Introduction}

Power consumption is among the largest expenses in data centers and is estimated at 1-1.5\% of global electricity use~\cite{masanet2020recalibrating}.
The recent surge in training Large Language Models (LLMs), consuming around 29.3 terawatt-hours per year—equivalent to Ireland's energy consumption~\cite{devries2023growing}—has prompted companies like Amazon, Google, and Microsoft to invest billions in nuclear energy~\cite{newyorktimes2024} to meet this demand.
The Frontier supercomputer, the world's first exascale supercomputer, consumes 22.7\,MW continuously~\cite{top500}.
Supercomputing was found to be responsible for 59.8\% of the carbon emissions of the average astronomer in Australia, 3.6 times as much as air travel~\cite{stevens2020imperative}.
As less than 15\% of the world’s energy comes from renewable sources~\cite{owid-renewable-energy}, it is crucial that we investigate how to improve the energy efficiency of these systems and applications that run on them, and reduce our carbon footprint.

Over the past decade, large improvements in the energy efficiency of data center cooling and power provisioning have been significant enough to nearly offset the growth of IT device energy use~\cite{masanet2020recalibrating,shehabi2018data}.
As such, the crucial next step is to understand and improve energy expenditure within computer systems.
Modern systems rely on many peripheral devices, including network interface controllers (NICs) and solid state drives (SSDs) that all require power.
Among the peripheral devices, Graphics Processing Units (GPUs) stand out as the primary computing platform for nearly all large-scale AI and HPC applications~\cite{lecun2015deep,heldens2020landscape}, delivering 99\% of the compute performance in modern supercomputers~\cite{frontier}, and consuming $>$64\% of the total power of these systems~\cite{stachowski2021autotuning}.
To advance energy efficiency research, it is critically important to develop fast, accurate, and openly accessible methods for measuring the power consumption of computer components.

Many software-based methods have been developed to effectively reduce energy consumption in computing systems. For example,
dynamic voltage and frequency scaling (DVFS)~\cite{ge2013effects,mei2013measurement,price2016optimizing,akiki2018energy},
power-aware scheduling~\cite{katagiri2013energy,guerreiro2015multi},
power capping~\cite{krzywaniak2019performance}, and
energy-efficient algorithm design~\cite{datta2008stencil,huang2009energy,dong2014step,schoonhoven2022going} can significantly lower power usage without compromising performance.
For instance, power-aware scheduling algorithms can optimally divide work between CPU and GPU depending on specific task properties~\cite{katagiri2013energy,li2015meterpu,garzon2017approach}, or generate optimized schedules for GPU kernels executing concurrently~\cite{guerreiro2015multi}. 
Additionally, code-level optimizations, such as compiler~\cite{nobre2018compiler, pallister2015identifying} and function-level~\cite{stachowski2021autotuning, schoonhoven2022going} tuning, can lead to substantial energy savings.
However, implementing these methods effectively requires fast (sub-millisecond) and accurate power measurements at a fine-grained level, for example down to single GPU kernels or even during the execution of individual operations.
Without precise measurement tools, it is challenging to assess the impact of optimizations and guide further improvements in energy efficiency.

This paper presents PowerSensor3, a tool that measures the instantaneous power consumption of SoC development boards and PCIe cards like GPUs, FPGAs, domain-specific accelerators for AI and ML, and NICs, at 20\,kHz (sub-millisecond) time scale.
PowerSensor3 includes several important improvements over PowerSensor2~\conditionalCite{romein2018powersensor} and other non-commercial and commercially available power measurement tools, including:
\begin{itemize}
%\item A modular design with a base PCB that supports up to 4 different sensor boards. %$ to accomodate different sensors.
\item A modular design with a base board that supports up to 4 sensor board modules. %$ to accomodate different sensors.
\item A variety of sensor boards, with different connectors and sensors (e.g., 8-pin PCIe power, USB-C, high-current and low-current boards with terminal blocks).
\item Support for measuring both voltages and currents.
\item An increased sampling rate from 2.8\,kHz to 20\,kHz through the use of a faster microcontroller. %using the more powerful STM32F401 ARM-based microcontroller
\item The use of current sensors that are hardly sensitive to changes of the external magnetic field.
\item Simplified, one-time calibration procedure through a command-line utility.
\item The base board and sensor boards are released as open hardware \cite{PS3-hw} (CERN-OHL-P v2) and the firmware and host library are released as open-source software \conditionalCite{PS3-sw} (Apache-2.0).
\item Cost-efficient design, a complete PowerSensor3 with 3 sensor boards costs less than €\,100 in components.
\end{itemize}

% Ben: In case we need to save space, the following paragraph is optional in my opinion
This paper is structured as follows.
\cref{sec:background} provides background and discusses related work.
\cref{sec:design-and-implementation} describes the PowerSensor3 design and implementation.
\cref{sec:evaluation} characterizes its performance, and in \cref{sec:application-case-studies}, we describe some application use cases.
Finally, \cref{sec:discussion} discusses the application and extendibility and \cref{sec:conclusions} concludes.
\section{Background and Related Work}\label{sec:background}

Various methodologies exist for measuring power consumption within computer systems. This section provides a comprehensive overview of the power measurement tools that have been utilized and discussed in the scientific literature.

Several researchers have used commercially available tools to measure whole system power, seeking to improve the energy efficiency of GPU applications.
However, these tools generally have very low sampling rates, for example
the Watts Up Pro operates at 1\,Hz~\cite{huang2009energy,ge2013effects,mei2013measurement,jia2015gpu}, 
Cray PMDB at 10\,Hz~\cite{anzt2015experiences}, or 
Yokogawa WT230 at 10\,Hz~\cite{grasso2014energy,schiffmann2018optimizing}. 
While measuring whole system power might give a realistic view of power consumption for the whole application, measuring power consumption of components in isolation can give insights needed to improve efficiency in critical parts of the application.

For CPUs, several software packages exist to monitor power consumption by reading from built-in sensors.
Intel’s Running Average Power Limit (RAPL) provides a number of performance counters to read energy consumption of CPUs and DRAM, with a time frequency of 1\,kHz~\cite{khan2018rapl}.
LIKWID also provides the likwid-powermeter tool, which builds on top of RAPL and allows to measure power consumption on architectures from Intel, AMD, ARM, and IBM.

However, many PCIe devices such as SSDs, domain specific accelerators or NICs do not have built-in power measurement tools, and thus require external power measurement.
And as such researchers have been developing their own custom-built power measurement devices for measuring power of such components~\cite{romein2018powersensor}.
One challenge with power measurement of PCIe devices is that PCIe devices receive power from different sources. Up to 75\,W of power could be delivered via the PCIe slot, 10\,W of which via the 3.3\,V rail, the rest at 12\,V. Devices that need more than 75\,W can receive additional power at 12\,V from the power supply unit through, possibly multiple, 6-pin, 8-pin or 12-pin PCIe connectors, or from the host motherboard using the 8-pin EPS connector.
As such, measuring the power consumption of PCIe devices requires measuring current across multiple power cables. Moreover, voltages cannot be assumed to be stable under load, therefore the voltage needs to be measured for every power cable as well.

\subsection{GPU on-board power measurement}

NVIDIA has been shipping an internal power sensor in both server-grade and consumer-grade GPUs as part of the Kepler architecture~\cite{nvidia2012kepler}, starting with the NVIDIA Tesla K20. 
While the properties of NVIDIA's current sensor have been studied widely~\cite{lang2013high,burtscher2014measuring,yang2024accurate}, only a few studies have used AMD's built-in current sensor and reported on its accuracy and sampling frequency. Wu et al.~\cite{wu2015gpgpu} reported that 
the AMD Radeon HD 7970 estimates the chip-wide dynamic power and updates
the power estimates every millisecond.
Schieffer et al.~\cite{schieffer2024rise} report to have used a sampling frequency as low as 10\,ms on the AMD MI250X GPU, but did not state the lower bound.

An evident advantage of utilizing these built-in sensors is their widespread availability. However, there are two significant drawbacks.
Firstly, a persistent issue with vendor-based APIs for reading the GPU internal power sensor is that they typically return only averaged power consumption values~\cite{burtscher2014measuring}. Notably, NVIDIA has addressed this with driver update 530 (May 2023), which extends their API to support instantaneous power readings~\cite{yang2024accurate}.
Secondly, the use of on-board power measurement in GPUs is hindered by low sampling frequencies. Even with the capability for instantaneous power readings, new values are provided at a frequency of approximately 10 Hz on NVIDIA GPUs~\cite{yang2024accurate}.

The issues with on-board power measurement in GPUs have forced researchers in GPU energy efficiency to artificially increase the execution time of their GPU kernels by several orders of magnitude to allow for enough samples to be collected and to overcome the effects of averaged power readings and low sampling rates~\cite{krzywaniak2019performance,price2016optimizing,chen2015angel,schoonhoven2022going}. There are several downsides to such approaches, as evaluating power consumption in different settings or for different software implementations consumes large amounts of time and energy, and the measured execution itself can be less realistic compared to real execution scenarios.

\subsection{External power measurement tools}

Several researchers have used current clamps to determine GPU power consumption~\cite{ren2011algorithm, katagiri2013energy, suda2013mathematical} without documenting the achieved accuracy and sampling rate. Timm et al.~\cite{timm2012design} also used a current clamp to determine GPU power consumption and mentioned a sampling rate of 10~kHz.

PowerMon2~\cite{bedard2010powermon} is a custom-built power monitoring device for voltage and current measurements with a 1\,kHz sampling frequency and a relatively low, -6.6\% / +6.8\%, current measurement accuracy. It also uses a difficult to obtain custom implementation and cannot handle 150\,W PCIe power cables. 
Another tool, PowerInsight~\cite{laros2013powerinsight}, measures both voltages and currents, but has a sample rate of less than 1\,kHz. Consequently, it cannot capture the detail required to conduct precise PCIe power measurements. The exact sample rate is not given in the paper.
Finally, NVIDIA has also produced the power measurement device Power Capture Analysis Tool (PCAT), which is not for sale nor is its design documented. The PCAT documentation suggests a sampling rate of 10\,Hz\footnote{https://developer.nvidia.com/nvidia-power-capture-analysis-tool}.

In addition to research-oriented PCIe power sensors, several commercial options are available. Most prominent are PMD (\$\,60) and Powenetics V2 (€\,975). The Powenetics V2 is expensive and has a sampling rate of up to 1\,kHz according to their website. PMD was recently used by Yang et al.~\cite{yang2024accurate} to perform a large study of the accuracy of power measurements reported by NVML on over 70 different GPUs. They mention that while PMD has an internal sampling frequency of 34\,kHz, PMD's (Windows-only) host library limits updates to a sampling frequency of 10\,Hz. Yang et al.~\cite{yang2024accurate} developed their own data logger to achieve a sampling frequency of 5\,kHz. To the best of our knowledge, their data logger is not openly available.

\section{Design and Implementation}
\label{sec:design-and-implementation}

To measure power for modern PCIe accelerators, it is essential to monitor both PCIe slot power (3.3\,V and 12\,V, up to 75\,W) and up to two external power connections (up to 600\,W) for PCIe gen 5 and 6. Accurate measurement requires monitoring voltage and current for each power supply with isolation to prevent coupling between the device and the sensors.
Modern accelerators, like GPUs, need fast sensors due to their high processing speeds and low kernel execution times. In order to reduce power loss over cables, the sensor must be close to the accelerator, installed within the server. This imposes size and safety constraints, requiring stable connectors and ensuring no contact with other components in a noisy environment.
The PowerSensor3 is designed to meet the needs of PCIe gen 4, 5, and 6 accelerators. Its flexible design makes it suitable for high-power PCIe devices (e.g., GPUs) as well as lower-power or standalone devices such as SoC boards.

\subsection{Hardware Design}

The core of the PowerSensor3 system is a baseboard that accommodates the ``Black Pill'' STM32F411 microcontroller module and up to four sensor modules. The STM32F411~\cite{datasheet-STM32F411} was selected due to its ability to sample up to sixteen analog inputs, enabling the support of four sensor modules, its USB data transfer capabilities, and the availability of software development tools. To cater to different power ranges and connectivity, several different sensor modules are developed. The designs are open-source and can be modified to meet a specific power range, accuracy or connector type. This modular approach allows users to select the most appropriate power monitoring sensors for their specific applications. Additionally, a small display is integrated into the baseboard to show instantaneous power consumption.

\begin{figure}[ht]
\centering
\includegraphics[width=0.5\textwidth]{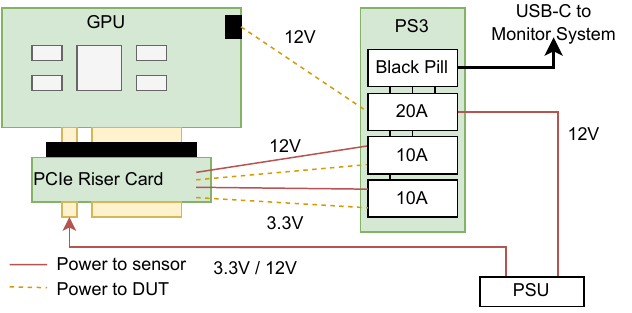}
\caption{Schematic of a PowerSensor3 measurement setup.}
\label{fig:PS3_Schematic_Overview}
\end{figure}

\cref{fig:PS3_Schematic_Overview} illustrates an example of the PowerSensor3 in operation. In this example, the PowerSensor3 is equipped with a PCIe sensor module that measures the power supplied to the PCIe card via the external power input as well as two sensor modules to measure the 3.3\,V and 12\,V PCIe slot power.
By utilizing a modified riser card, where the power connections for both 3.3\,V and 12\,V are interrupted and routed through two sensor modules, it is possible to measure the power consumption of both the slot and the external connection without affecting the PCIe signal integrity.

Each sensor module measures both voltage and current. To mitigate the effect of ground loops on the measurements, the circuit connected to the Device Under Test (DUT) is isolated from the measurement logic connected to the microcontroller. This isolation is achieved using a differential Hall sensor, the Melexis MLX91221~\cite{datasheet-MLX91221}, for current measurement, and an optically isolated voltage sensor, Broadcom ACPL-C87B~\cite{datasheet-ACPL-C87B}, for voltage measurement. The Hall sensor family supports a variety of pin compatible devices with different current ranges.

The resistance loss in both the power and return paths of the power sensor, which can cause measurement inaccuracies and affect the power delivered to the DUT, is a critical design parameter. To minimize resistance, the sensor is designed to be compact.
To reduce the impact of voltage loss within the sensor and the connecting wire to the DUT, a remote sense connector is integrated into the sensor module. This allows for the measurement of voltage directly at the DUT rather than at the input port.

\begin{figure}[h]
\centering
\includegraphics[width=0.45\textwidth]{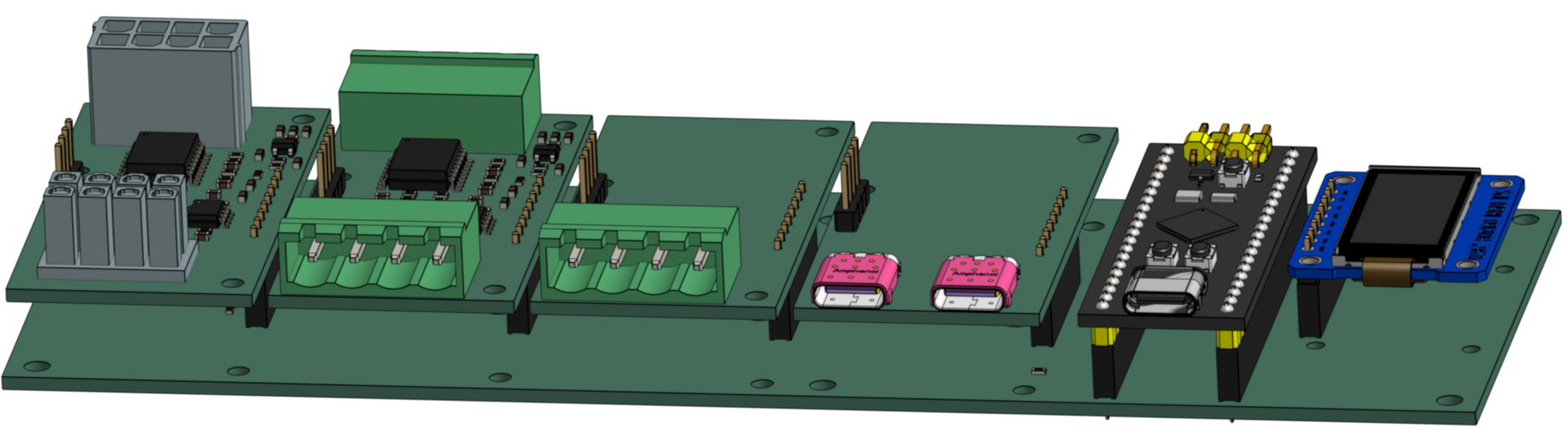}
\caption{3D rendering of PowerSensor3 with PCIe 8-pin, 20\,A, 10\,A, USB-C sensor modules, ``Black Pill'' module and display.}
\label{fig:PS3_3D_render}
\end{figure}

PowerSensor3 currently comes with five different designs for sensor modules:
\begin{itemize}
    \item 20\,A PCIe 8-pin: With a connector for easy integration with the external power connector on PCIe cards.
    \item 10\,A: Designed to measure power between the PCIe slot and the PCIe card.
    \item USB-C: Suitable for USB-powered systems.
    \item 20\,A: General-purpose power measurement for medium power applications, with terminal block connectors.
    \item 50\,A High-Current: For high-power applications.
\end{itemize}
These sensor modules can be combined in various configurations within a single setup, providing a comprehensive and adaptable power measurement solution. \cref{fig:PS3_3D_render} shows a 3D rendering of a populated PowerSensor3 module. The design of the baseboard and sensor modules has been made available \conditionalCite{PS3-hw}.

%\textcolor{red}{[link]}

The analog signals from the sensor modules are passed to the STM32F411, where the 10 most significant bits of the Analog to Digital Converter (ADC) are utilized. The bandwidth of the current and voltage sensors is well above the ADC's output sample rate of 20\,kSamples/s. The maximum time resolution of the current sensor is specified at 300\,kHz, while the maximum time resolution of the voltage sensor is 100\,kHz.

The measured power is calculated with:
\begin{equation*}
P = (U + E_u) \cdot (I + E_i)
\end{equation*}
where \( U + E_u \) is the combination of the voltage and the error in the voltage reading, and \( I + E_i \) is the combination of the current and the error in the current reading. This gives the error in the power reading as:
\begin{equation*}
E_p = \sqrt{(U \cdot E_i)^2 + (I \cdot E_u)^2 + (E_i \cdot E_u)^2}
\end{equation*}
This formula shows that for small loads, the noise is dominated by the error in the current reading, while for low voltage high current sensors, the noise is dominated by the error in the voltage reading.

The error in the current reading ($E_i$) consists of the error due to the quantization noise of the ADC, combined with the sensitivity of the Hall sensor and inherent noise of the Hall sensor. Among these factors, the noise in the Hall sensor, which is 115\,mArms for the 10\,A sensor, is the dominant factor, resulting in a peak-to-peak error of 4.1\,Wpp.

The error in the voltage reading ($E_u$) is caused by quantization noise and the inherent amplifier noise. The noise on the voltage reading is increased due to the voltage divider. For a 12\,V / 10\,A sensor module, the noise on the voltage at high currents is estimated to be 0.2\,Wpp.

\cref{tab:ps3_module_accuracy} provides an overview of the theoretical worst case accuracy of the sensor modules.

\begin{table}[tb]
    \centering
    \caption{Theoretical worst case accuracy of PowerSensor3 modules.}
    \label{tab:ps3_module_accuracy}
    \begin{tabular}{l | c c c} 
       
        \textbf{Module} & \textbf{Voltage} & \textbf{Current} & \textbf{Power} \\ 
        \hline
        12\,V / 10\,A  & $\pm$ 28.6\,mV & $\pm$ 0.35\,A & $\pm$ 4.2\,W \\ 
        3.3\,V / 10\,A & $\pm$ 19.9\,mV & $\pm$ 0.35\,A & $\pm$ 1.2\,W \\ 
        USB-C (20\,V / 10\,A) & $\pm$ 28.6\,mV & $\pm$ 0.35\,A & $\pm$ 7.0\,W \\ 
        Ext (12\,V / 20\,A) & $\pm$ 28.6\,mV & $\pm$ 0.41\,A & $\pm$ 5.0\,W \\ 
       
    \end{tabular}
\end{table}

\subsection{Firmware Design}
This section details the firmware design using the STM32F411 microcontroller. Its primary function is to read current and voltage sensors at a high, constant rate and transmit the data to the host via USB. The ADC continuously reads the sensors, and the DMA controller transfers the values to RAM. Once all sensor data is in RAM, an interrupt is generated. The interrupt handler reads the sensor values, adds metadata, and creates a data package for the host. A main loop checks for data to be sent and transmits it as needed.

We use the STM32 low-level library through \emph{STM32duino}\footnote{\url{https://github.com/stm32duino}}. \emph{STM32duino} provides a simple interface, integrating it into the Arduino ecosystem, and as such facilitating firmware development, compilation, and uploading through the widely-adopted Arduino tools.

Ideally, the ADC would operate at the highest possible clock speed, transmitting data directly to the host. However, the data rate would exceed the capacity of the USB controller on the Black Pill, which supports up to USB 1.1 full speed (12\,Mbit/s). Although a USB 2.0 controller can be added, we opt to reduce the sampling rate instead, to minimize the cost and complexity of the PowerSensor3 hardware.

Each sensor board contains a sensor pair with current and voltage sensors. Within each pair the sensors are connected to consecutive ADC channels, minimizing the time difference between measurements. The ADC operates at a clock speed of 24\,MHz, with the CPU averaging several samples to reduce the final sampling rate.
The ADC is configured with a 10-bit resolution and a sampling time of 15 clock cycles. Each bit requires one cycle to read, resulting in a total ADC sampling time of 25 clock cycles or 1.04\,$\mu s$. Reading 8 sensors (4 modules) and averaging 6 consecutive samples on the CPU amounts to a 50\,$\mu s$ interval, corresponding to a sampling rate of 20\,kHz.
For each sensor, we transmit 2 bytes of data to the host. With 10 bits per sensor value and 6 bits for metadata: the sensor index, a marker, and one bit in each byte to differentiate the first byte from the second.

The sensor data sent from the PowerSensor3 to the host is preceded by a device timestamp. The timestamp is generated after processing 3 out of the 6 samples to be averaged and is stored as a 10-bit value in microseconds.
Since there is no room in the sensor data packets for the timestamp, it is sent separately. To differentiate the timestamp data from sensor data, a combination of the sensor and marker bits is used: a real marker bit can only be set in the sensor data of sensor 0. A marker bit set to one with a nonzero sensor index is unused and can be repurposed for other data. For the timestamp, the maximum sensor index of 7 (binary 111) is used.

The firmware supports several options through the host:
\begin{itemize}
    \item Start or stop streaming of sensor data.
    \item Send or receive configuration values (\cref{sec:fw_sensor_config}).
    \item Send a marker with the next sensor data.
    \item Send the firmware version as a string.
    \item Reboot the device, optionally to DFU mode which is used for uploading new firmware.
\end{itemize}

\subsubsection{Sensor configuration values}
\label{sec:fw_sensor_config}
The PowerSensor3 firmware is generally agnostic to the type of sensor module used. However, the host software must know how to convert raw sensor values to accurate voltage and current readings. These conversion values are stored on the device and communicated to the host library, so the user does not need to keep track of the specific sensors used. The STM32 supports a virtual EEPROM implementation that stores data in flash memory. The following data is stored for each sensor:
\begin{itemize}
    \item Sensor name.
    \item Reference voltage.
    \item Sensitivity (current sensors) or gain (voltage sensors).
    \item Sensor state (enabled or disabled).
\end{itemize}

\subsubsection{Display}
PowerSensor3 is equipped with a compact display for real-time visualization of sensor values when the sensor is not in use by the host system. This display prominently features the total power consumption, while individual current, voltage, and power measurements for each sensor pair are shown in smaller fonts.

The display is connected through an SPI interface, controlled by the open-source Adafruit ST7735 library\footnote{\url{https://github.com/adafruit/Adafruit-ST7735-Library}}. To enhance display update speed and reduce CPU load, we expanded the library with two features: 1) we enable DMA for transferring the display buffer from RAM to the SPI controller, and 2) 
we pre-compute graphics for all necessary characters in all used color and size combinations, storing the resulting fonts in the program memory. A Python script for automatic font generation is included in the PowerSensor3 repository.

\subsection{Software Design}

The PowerSensor3 host library is implemented in C++, with an optional Python wrapper\footnote{implemented using pybind11, \url{https://github.com/pybind/pybind11}}. The device is accessed via a PowerSensor C++ class, which, upon initialization, connects to the microcontroller and reads the sensor configuration values. Methods are available to read or set these values. A lightweight thread continuously receives sensor values from the device, and the library internally tracks the cumulative energy consumption measured by each sensor.

PowerSensor3 can operate in two modes: interval-based and continuous, both of which can be active simultaneously.
In interval-based mode, the user requests sensor states at two different times to calculate total energy consumption and average power. This mode can be accessed through a standalone executable or via the C++ or Python interface, allowing precise control over the measurement period but requiring source code modification. The standalone executable, \emph{psrun}, connects to PowerSensor3, runs the provided application (executable), and reports the total energy consumed after execution.
In continuous mode, PowerSensor3 records all sensor data to a file at 20\,kHz resolution. The library supports custom marker characters in the output file, time-synced with the microcontroller, to correlate timestamps with specific parts of the application code.

The PowerSensor3 host library comes with three additional executables for easier interfacing with the device:

\emph{psconfig} reads or writes the sensor configuration values and optionally reboots the device. After installing the firmware, this tool is used to configure the device.

\emph{psinfo} shows the configuration values of each enabled sensor, as well as the latest measurements and the total power.

\emph{pstest} measures and reports power and energy at increasing intervals for testing purposes.

\subsection{Calibration}

The sensor modules are calibrated using a known power supply, such as the system's power supply unit or a laboratory power supply. During calibration, the sensor modules are unloaded (no power dissipation), and the voltages on the voltage sensors are measured. By taking 128\,k samples and calculating the average current and voltage readings, the offset error of the Hall current sensor and the gain error for the voltage are determined. These corrections are then stored in the microcontroller. Python scripts are available to guide through this process. Based on the measurements described in \cref{{sec:evaluation}}, calibration is only required once at production. 
\section{Evaluation}
\label{sec:evaluation}
%In this section:
%- Accuracy of sensors, what is the error? (relative / absolute)
%- Noise measurements?
%- Voltage drop measurements
%- direct v.s. remote sensing
%- Temperature measurements
%- Long term stability (no need to re-calibrate)

\begin{figure}[t]

\includegraphics[width=0.45\textwidth]{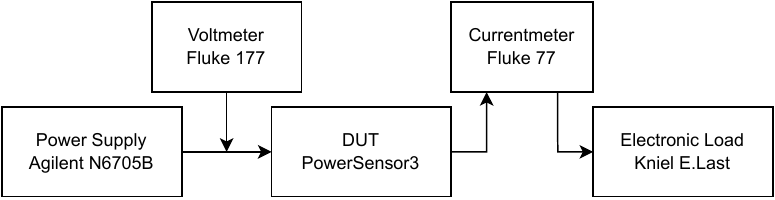}
\caption{Measurement setup for accuracy assessment.}

\label{fig:PS3_measurement_setup}
\end{figure}

% text before the update
To verify the functionality of the PowerSensor3, the test setup illustrated in \cref{fig:PS3_measurement_setup} was utilized. A laboratory power supply (Keysight~N6705B) served as the power source for the DUT. An electronic load (Kniel E.Last) was employed for loads up to 10\,A. The voltage at the sensor and current through the load were measured using a Digital Multimeter (Fluke~177 for the voltages and Fluke~77 for the current). Data was captured using \emph{pstest}.

\subsection{Sensor accuracy assessment}

To evaluate the sensor's accuracy, a measurement was conducted where the load current was swept in 1\,A steps from the minimum (-10\,A) to the maximum current (+10\,A). At each step, 128\,k samples were collected using the \emph{pstest} tool.
This data allowed the determination of the accuracy and variability of the current and voltage readings, and thus the power calculation.

\begin{figure}[]

\includegraphics[width=0.5\textwidth]{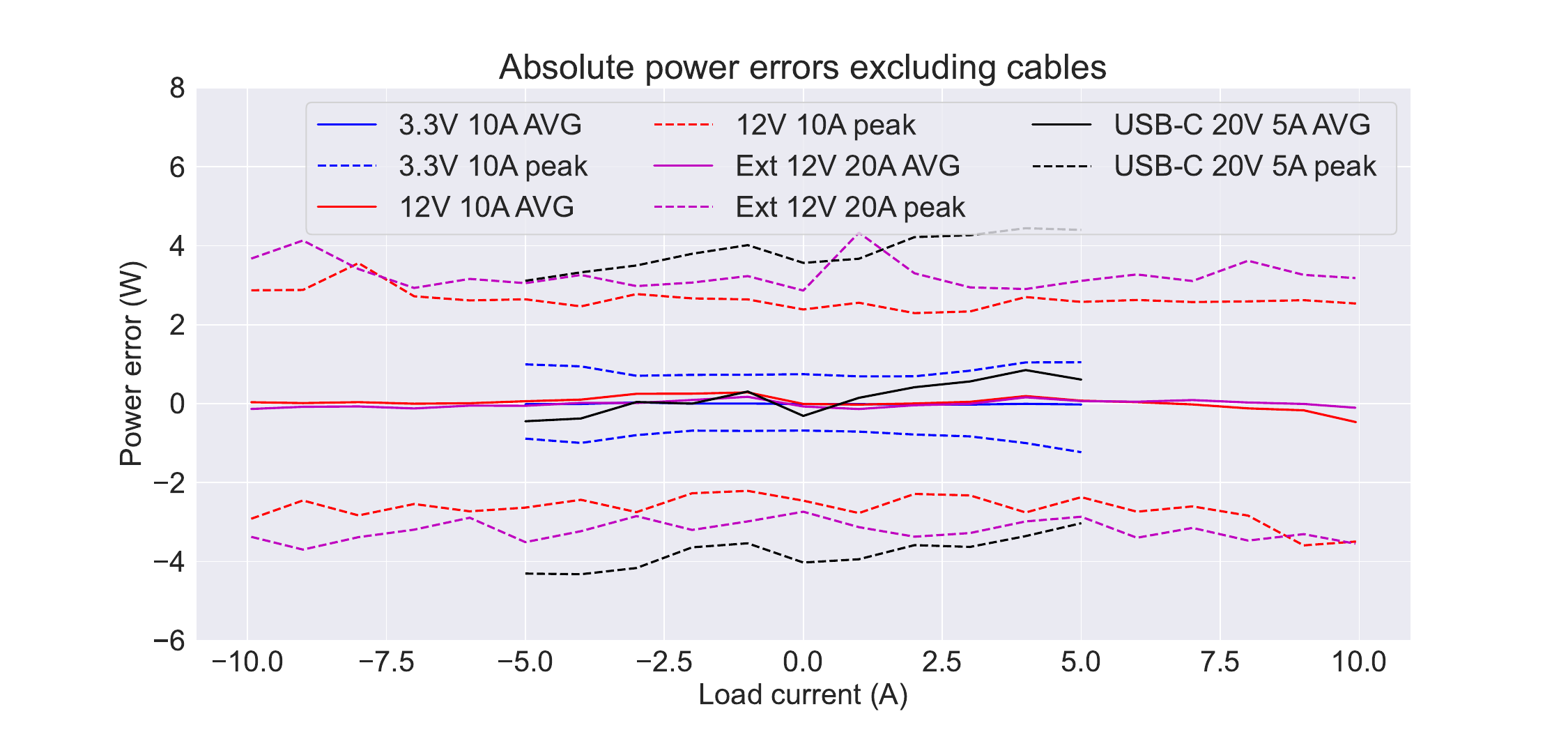}
\caption{Power error for four types of sensors with dotted lines indicating min and max values per measurement point.}

\label{fig:PS3_current_error}
\end{figure}

In \cref{fig:PS3_current_error}, the results are shown. The continuous line indicates the difference between the expected power and the measured power. The dotted lines in this figure represent the minimum and maximum difference within the 128\,k samples at each measurement point. As can be seen in this figure, the accuracy of the 3.3\,V sensor is better in comparison with the 12\,V sensor, where the error in the current sensor is multiplied by 12 instead of 3.3.

Detailed inspection of the data indicates that at low currents, noise originates primarily from the current sensor, while at higher currents, the voltage sensor noise becomes more significant. Averaging the samples can reduce the noise but also lowers the time resolution ($F_s$). \cref{table:averaging_effect} provides an error overview for a 12 V, 10 A sensor with an 8\,A load, where blocks of samples are averaged. In the table, the minimum and maximum values after averaging, the peak-to-peak range between these two values, and the standard deviation are shown.

\begin{table}[]
\label{tab:ps3_module_measured_accuracy}
\scriptsize
\centering
\caption{Overview of error values for different samples rates for 0.5\,A and 1\,A loads.}
\label{table:averaging_effect}
\begin{tabular}{c|cccc|cccc}
\multicolumn{1}{c|}{$F_s$} & \multicolumn{4}{c|}{0.5 A load} & \multicolumn{4}{c}{1 A load} \\ 
 & min & max & p-p & std & min & max & p-p &  std \\ 
 kHz & $W$ & $W$ &$W_{pp}$&$W_{rms}$& $W$ & $W$ &$W_{pp}$& $W_{rms}$\\ \hline
20 & 2.78 & 9.16 & 6.381 & 0.718 & 7.79 & 15.48 & 7.685 & 0.722 \\ 
10 & 4.04 & 8.22 & 4.173 & 0.507 & 9.42 & 14.53 & 5.109 & 0.511 \\ 
5 & 4.85 & 7.69 & 2.842 & 0.358 & 10.54 & 13.68 & 3.142 & 0.362 \\ 
1 & 5.66 & 6.85 & 1.183 & 0.16 & 11.62 & 12.9 & 1.285 & 0.163 \\ 
0.5 & 5.85 & 6.67 & 0.821 & 0.113 & 11.92 & 12.73 & 0.814 & 0.117 \\ 
\end{tabular}
\end{table}

\subsection{Long term stability}

The long-term stability of PCIe 8-pin sensor modules was assessed using the setup in \cref{fig:PS3_measurement_setup} with a 7.5\,A load. Over 50 hours, 128\,k samples were taken every 15 minutes using \emph{pstest}. Average, minimum, and maximum power values were calculated for each point. Marginal fluctuations ($\pm$ 0.09\,W) were observed in the average values, with more noise in the minimum and maximum values. These results indicate that the PowerSensor3 remains stable and does not require recalibration after production.

\subsection{Step response}
\begin{figure}[tb]
\includegraphics[width=0.45\textwidth]{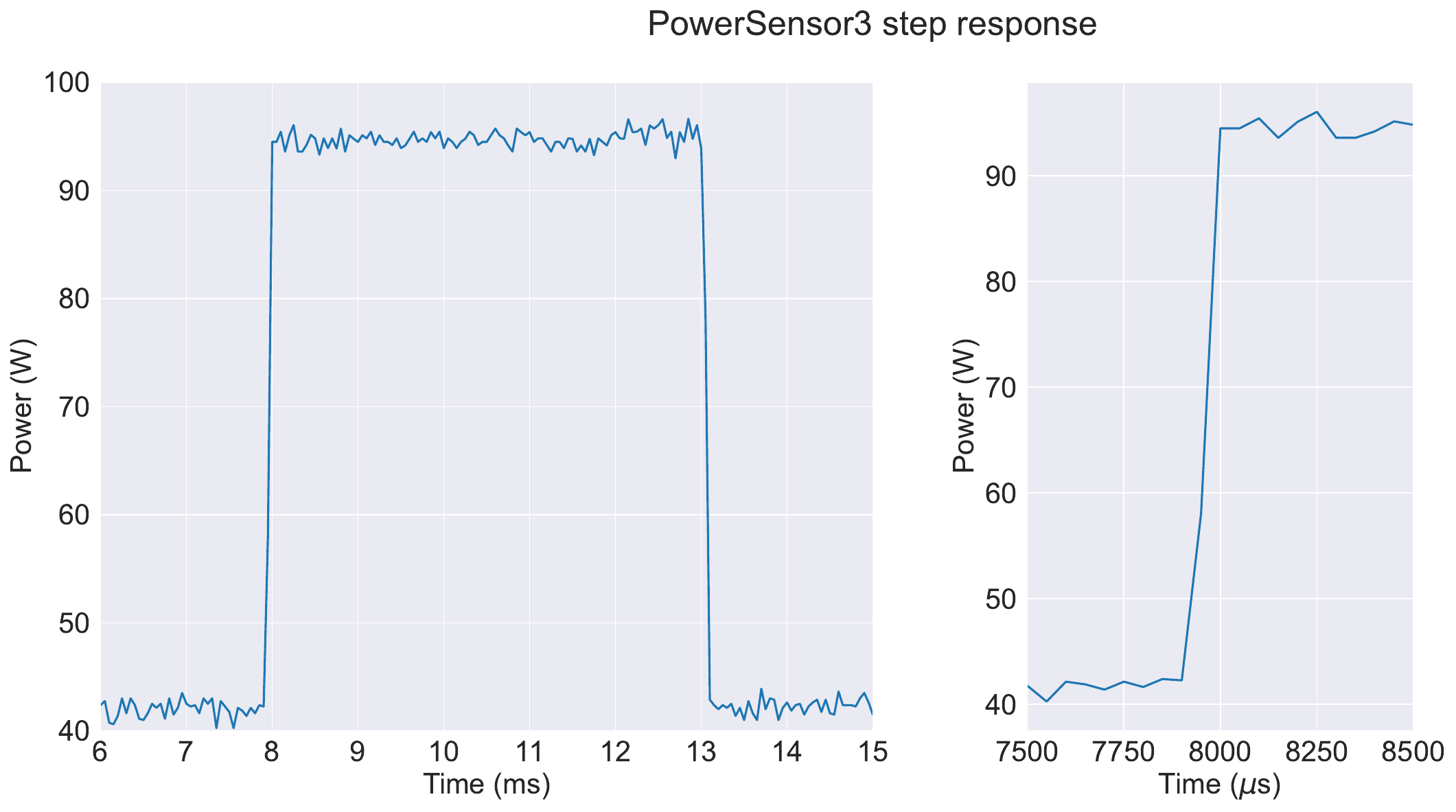}
\caption{Step response of PowerSensor3: load stepped from 3.3\,A to 8\,A plotted in ms scale (left) and $\mu$s scale (right).}
\label{fig:PS3_step}
\end{figure}

To measure the step response of the PowerSensor3, a 12\,V / 10\,A sensor, sampling at 20 kHz, is connected to the electronic load. The load is configured to 8\,A, with a frequency modulation of 100\,Hz and a modulation depth of 50\%. The results are shown in \cref{fig:PS3_step}. The step response is clearly visible, illustrating that the PowerSensor3 is well suited to measure power transients, such as the start and stop of a GPU kernel.

\section{Application case studies}
\label{sec:application-case-studies}

This section presents three use cases of PowerSensor3 demonstrating its capability to give highly-detailed insights into the power consumption of peripheral and embedded devices. The three case studies are: (1) discrete GPUs, (2) SoC boards, and (3) SSDs.

\subsection{GPUs}

\begin{figure}[t]
\centering
\ifdefined\DOUBLEBLIND
    \subfloat[PowerSensor3 attached to an AMD W7700 GPU.]{\includegraphics[width=0.45\textwidth]
{figures/DAS6_node504_W7700_blind.JPG}}
\else
    \subfloat[PowerSensor3 attached to an AMD W7700 GPU.]{\includegraphics[width=0.45\textwidth]
{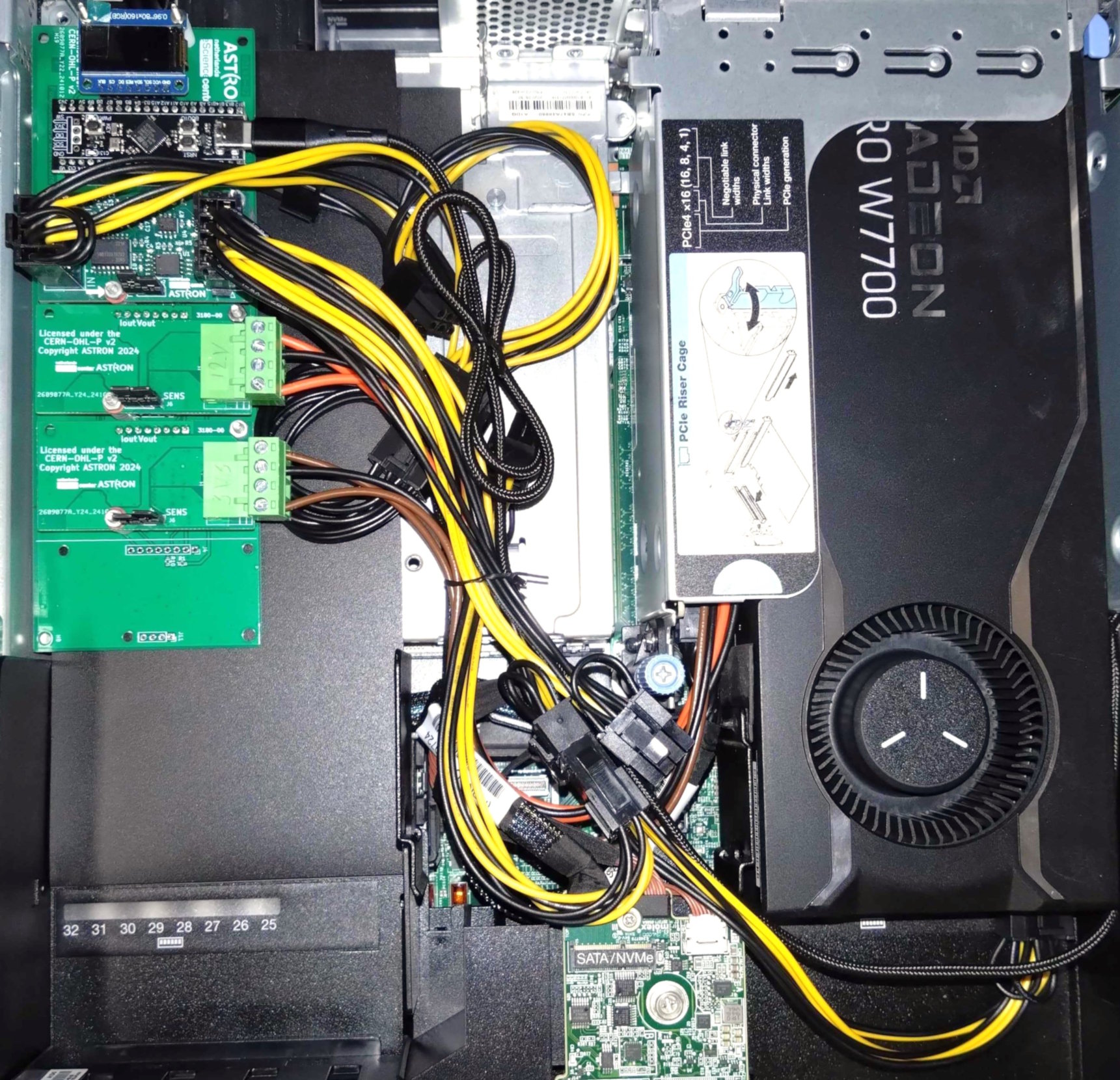}}
\fi

\subfloat[Modified PCIe gen 4 riser card (Lenovo SR665), providing measurement points for 3.3\,V and 12\,V.]{\includegraphics[width=0.45\textwidth,angle=180]{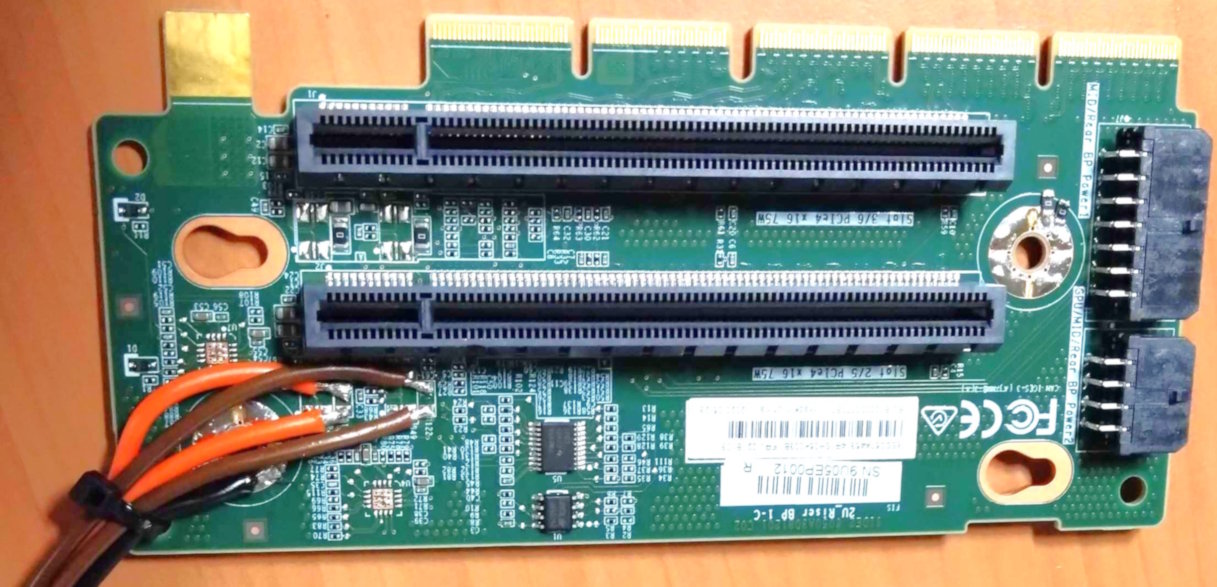}}%

\caption{A node in the DAS6 \conditionalCite{DAS6} cluster computer equipped with a PowerSensor3 to measure GPU power consumption.}
\label{fig:DAS6}
\end{figure}

In this section, we illustrate the application of PowerSensor3 in monitoring the power consumption of GPU applications. We have equipped multiple compute nodes in the DAS6 cluster \conditionalCite{DAS6} with a PowerSensor3 with 3 sensor boards as shown in \cref{fig:DAS6}(a). Two sensor boards for the 3.3\,V and 12\,V PCIe power channels and one for the 12\,V PSU power channel.
The PCIe power channels are intercepted using a modified PCIe Gen 4 riser card as shown in \cref{fig:DAS6}(b). By removing a 0-ohm resistor and attaching wires for each power channel, we created measurement points for the power supplies without compromising signal integrity.
We consider two use cases. First, we use PowerSensor3 to monitor the power consumption of a single kernel executing on the GPU and compare the results with the GPU's internal power sensor using Power Measurement Toolkit. Secondly, we use PowerSensor3 to monitor the power consumption while automatically optimizing a realistic GPU application for both compute performance and energy efficiency using Kernel Tuner.

\subsubsection{Power Measurement Toolkit}
The Power Measurement Toolkit (PMT) is an open-source high-level software library for measuring and monitoring power consumption across various hardware platforms \conditionalCite{Corda2022}. Written in C++, PMT leverages vendor-specific APIs to collect power usage data. For NVIDIA GPUs, it uses NVML and, for AMD GPUs, ROCm SMI and its successor AMD SMI are both supported, while for CPUs, it utilizes the RAPL interface or LIKWID \cite{Treibig2010}. Additionally, PMT supports profiling other architectures, such as AMD FPGAs, and its modular design allows for straightforward extension to new hardware.

PMT provides a unified interface for power measurement, catering to both C++ and Python applications. PMT is particularly suited for high-performance computing researchers and developers, who can use it to evaluate and optimize energy efficiency, but it is equally valuable for a general-purpose public requiring a simple yet effective software-based power measurement tool.

Yang et al. examined over 70 GPUs across 12 architectural generations and revealed significant inaccuracies in NVML power readings, leading to severe under- or overestimates of energy consumption~\cite{yang2024accurate}. Although mitigations were proposed, these issues underline the need for more reliable alternatives. PMT addresses these concerns by offering support for PowerSensor3, which provides accurate and consistent power measurements without the caveats identified in NVML.

\begin{figure}[ht]
    \centering
    \subfloat[NVIDIA's NVML library provides two different energy  measurements, 'instantaneous' and 'average'. The former provides a better sampling rate. High time-resolution energy measurements such as PowerSensor3 uncover GPU behavior that is not visible with NVML.]{
        \includegraphics[width=\linewidth]{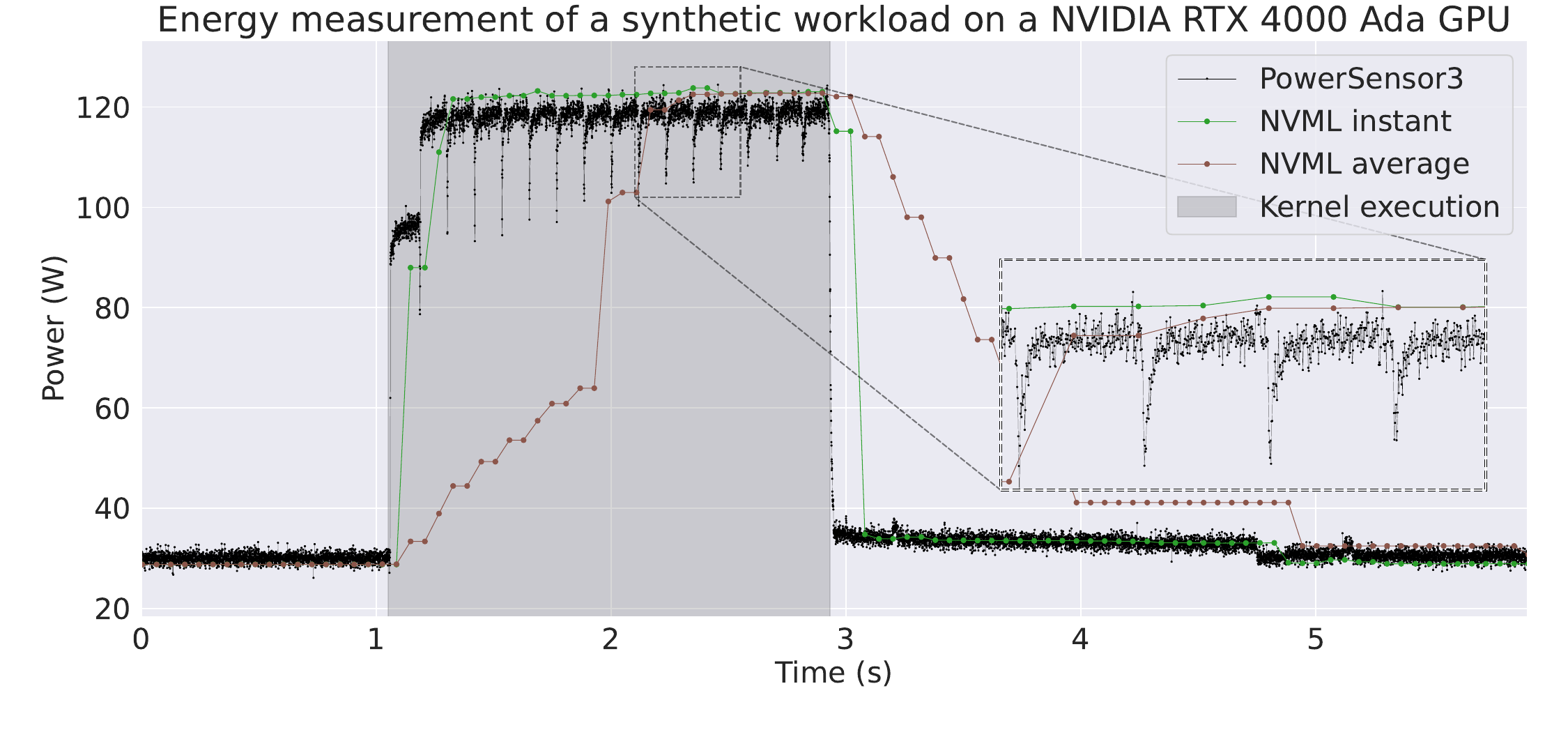}
        \label{fig:pmt-comparison-ad4000}}
    \newline
    \subfloat[The AMD SMI and PowerSensor3 measurement closely align.]{
        \includegraphics[width=\linewidth]{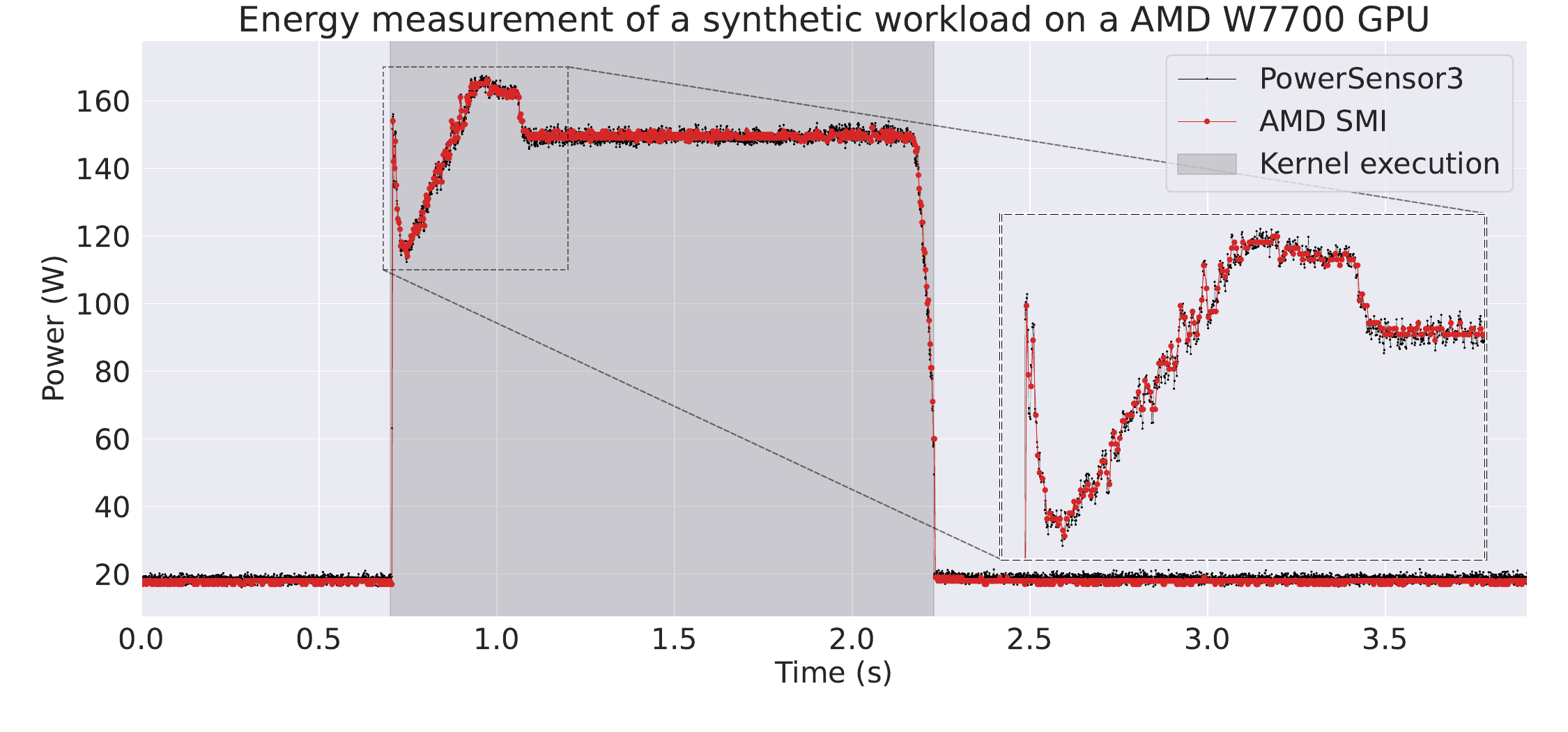}
        \label{fig:pmt-comparison-w7700}}
        \caption{Energy measurements for a synthetic GPU workload using PowerSensor3 and vendor supplied software-based measurements. The shaded area marks the kernel execution, and the insets highlight specific GPU behavior uncovered by energy analysis of the workload.}
        \label{fig:pmt-comparison}
\end{figure}

We compare the PowerSensor3 energy measurement with NVML on an NVIDIA RTX 4000 Ada GPU in \cref{fig:pmt-comparison-ad4000} and with AMD SMI on an AMD W7700 GPU in \cref{fig:pmt-comparison-w7700}. The measurement starts with a brief idle time, followed by a synthetic load of fused multiply-add instructions. A two-dimensional grid is used, where the x-dimension of the grid is set according to the number of streaming multiprocessors (SMs) or Compute Units (CU) on the NVIDIA and AMD GPU, respectively. The y-dimension is set such that the kernel runs for roughly two seconds. 

On the NVIDIA RTX 4000 Ada GPU, energy consumption initially spikes to approximately 95\,W before increasing to around 120\,W. This behavior corresponds to the gradual ramp-up of the clock frequency, which does not reach its peak instantaneously. Distinct phases are visible in the energy profile, corresponding to the sequential execution of thread blocks along the y-dimension of the grid. The power dips between individual phases are made clearly visible by PowerSensor3, but are missed by NVML. After the workload completes, the GPU requires over a second to return to its idle power state. While NVML’s instantaneous energy measurement aligns reasonably well with PowerSensor3, its time resolution cannot capture fine-grained GPU behavior. NVML's 'legacy' average power measurement is limited to coarse-grained energy estimations and completely inadequate to measure the kernel's energy use accurately.

On the AMD W7700 GPU, we compare PowerSensor3 with both ROCm SMI and AMD SMI APIs, which yields identical results despite differences in their programming interfaces. Unlike NVML, the built-in energy measurements of the W7700 GPU closely match PowerSensor3, demonstrating excellent accuracy. The energy profile reveals distinct phases of the GPU’s power and frequency behavior: an initial spike to the 150\,W power limit is followed by a sharp drop, a ramp-up phase with brief power overshoot, and eventual stabilization at the power limit. Notably, the GPU returns to its idle power state more rapidly than the NVIDIA GPU.

In conclusion, high-resolution energy measurement tools like PowerSensor3 are critical for uncovering GPU behavior that remains invisible to standard performance profilers, particularly in capturing transient power fluctuations that are not detectable at lower sampling rates. Given the limited disclosure from vendors regarding their built-in energy measurement mechanisms, reliable external tools are essential to ensure accurate and detailed energy analyses.

\subsubsection{Kernel Tuner}
\label{sec:kt}

We now use PowerSensor3 to provide power measurements while automatically optimizing a real-world GPU kernel for computational performance as well as energy efficiency. 

We use the Tensor-Core Beamformer~\cite{oostrum2025tcbf} as an example GPU application. Beamforming is a well-known technique to combine signals from multiple receivers. The Tensor-Core Beamformer has been developed for use in both radio astronomy and ultrasound imaging. The beamformer uses tensor cores on NVIDIA GPUs or matrix cores on AMD GPUs to perform complex matrix multiplications, which are not supported by vendor libraries such as cuBLAS or CUTLASS. In this case study, we use 16-bit input and output data with M=4096 beams, N=4096 samples at a time, and K=4096 elements summed.

The Tensor-Core Beamformer can be automatically tuned to achieve optimal performance on a specific GPU using Kernel Tuner~\cite{vanwerkhoven2019kernel}. Kernel Tuner is an open-source GPU auto-tuner that allows users to define parameters in the code to be tuned. The auto-tuner constructs a search space of all possible functionally-equivalent code variants and automatically searches for the specific combination of tunable parameter values that achieves the best performance. During the auto-tuning process, Kernel Tuner performs many empirical measurements to obtain the execution time and power consumption of each variant. In a typical use case, the tuner compiles and benchmarks several thousands of different code variants on the GPU.

Kernel Tuner supports capturing the energy consumption of GPU kernels~\cite{schoonhoven2022going}, which is typically measured using on-board current sensors, either using NVML on NVIDIA GPUs or ROCm-SMI through PMT for AMD GPUs. However, as explained in 
\cref{sec:background} and confirmed in \cref{fig:pmt-comparison-ad4000}, NVIDIA's on-board current sensors typically have a time resolution of about 10\,Hz, which is much too low to accurately capture the power consumed by real-world GPU kernels, which typically take at most a few tens of milliseconds. When using onboard current sensors for power measurement, Kernel Tuner therefore first executes the GPU kernel repeatedly to determine the execution time, and then runs the kernel continuously for an extended period, for example, 1 or 2 seconds, to collect sufficient measurements from the on-board sensor. As the tuner typically benchmarks several thousands of code variants, this means the tuning process is extended by several hours, which wastes both time and energy.

We have integrated support for PowerSensor3 directly into Kernel Tuner, which allows for instant capturing of the energy consumption of GPU kernels. In this way, there is no need for Kernel Tuner to continuously run the kernel for several seconds, effectively saving hours of tuning time. 

To auto-tune the Tensor-Core Beamformer for both energy and time efficiency on the NVIDIA RTX 4000 Ada GPU, we used the performance model presented in~\cite{schoonhoven2022going} to narrow down the range of GPU clock frequencies to tune for. The other tunable parameters that can be varied in the code are the thread block dimensions,
the number of submatrices (fragments) per thread block
and per warp,
and the extent to which double buffering in shared memory is applied.
In total there are 512 different code variants, with 10 different GPU clock frequencies, this amounts to an auto-tuning search space of 5120 configurations, that are averaged over 7 trials each.

\begin{figure}
\centering
\includegraphics[width=1.0\columnwidth]{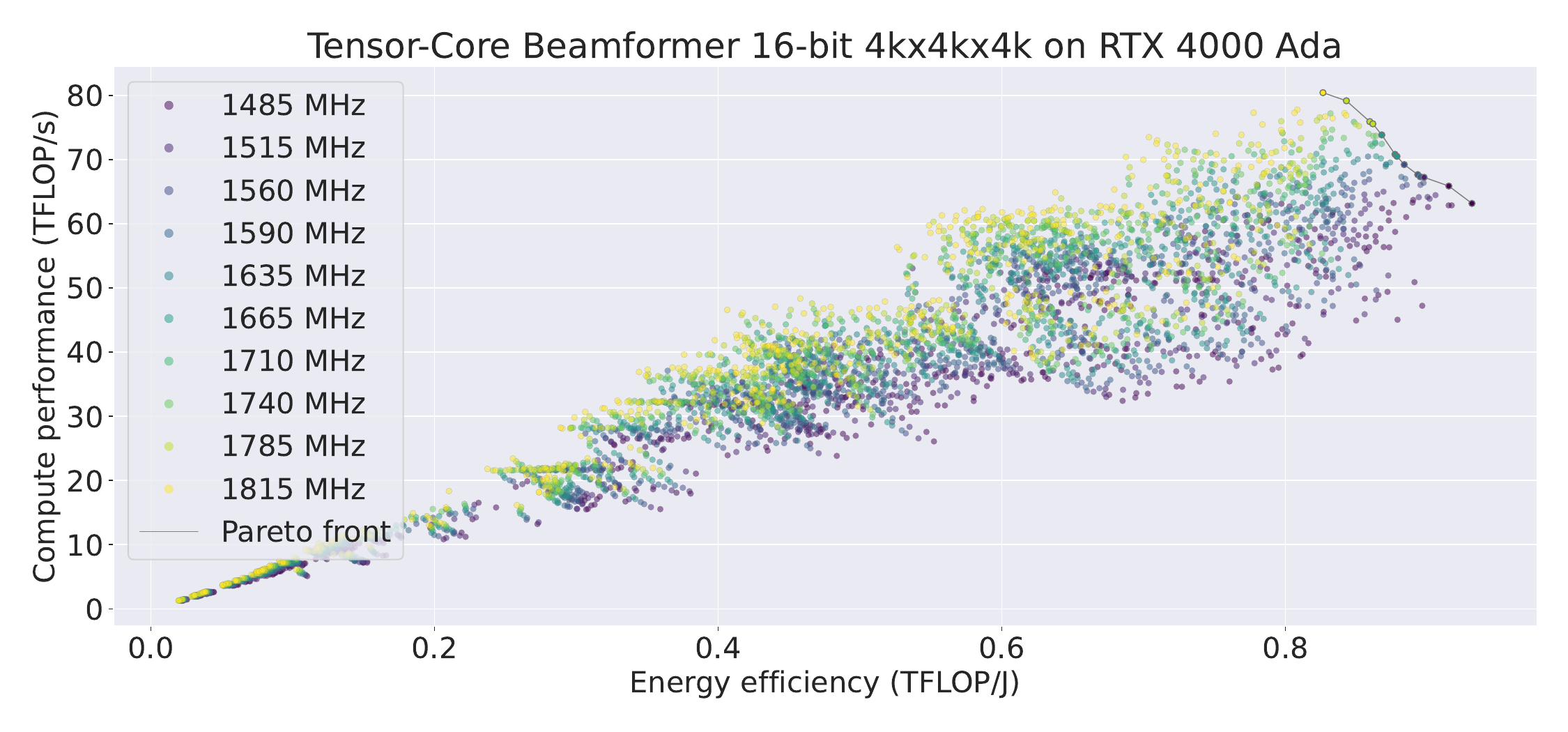}
\vspace{-0.7cm}
\caption{Tuning results for the Tensor-Core Beamformer on the NVIDIA RTX 4000 Ada.}\label{fig:beamformer-ad4000}
\end{figure}

\cref{fig:beamformer-ad4000} shows the energy efficiency in tera-flop per joule (TFLOP/J) and compute performance in tera-flops per second (TFLOP/s) of the code variants benchmarked during auto-tuning the Tensor-Core Beamformer on the NVIDIA RTX 4000 Ada.
Overall, we observe that performance and energy efficiency are correlated. However, especially among the more efficient configurations, there is a wider spread in both energy and compute efficiency.
The fastest Pareto optimal configuration achieves a compute performance of 80.4 TFLOP/s at 0.83 TFLOP/J energy efficiency, whereas the most energy efficient configuration is 12.7\% more energy efficient, but also has a 21.5\% slowdown compared to the fastest configuration.
Overall, collecting all data points from \cref{fig:beamformer-ad4000} using PowerSensor3 took 2274.4 seconds, which would have taken about 7394 seconds if we had used the onboard power sensor instead. Thus, thanks to PowerSensor3, we were able to perform this experiment in 3.25x less time.

\subsection{SoC boards (NVIDIA Jetson)}
\begin{figure}
\centering
\ifdefined\DOUBLEBLIND
    {\includegraphics[width=0\columnwidth]{figures/Jetson Orin labeled _blind.jpg}}
\else
 {\includegraphics[width=\columnwidth]{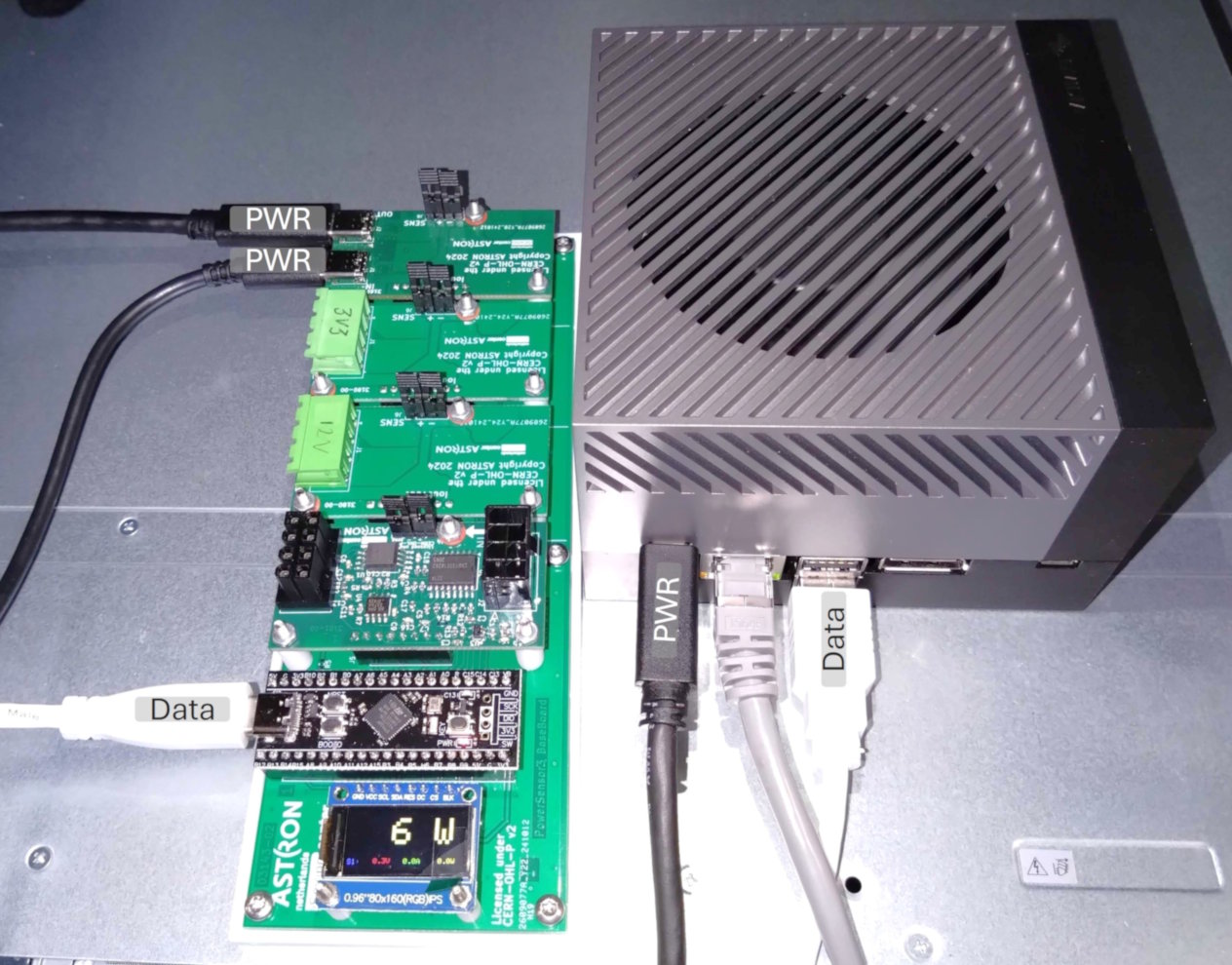}}
\fi
\caption{NVIDIA Jetson AGX Orin with PowerSensor3 on the USB-C power supply, the display shows the idle power.}\label{fig:jetson-orin}
\end{figure}

The NVIDIA Jetson series System-on-Chips contain a tightly integrated CPU and GPU, and are used in GPU-accelerated edge-computing systems.
\cref{fig:jetson-orin} shows an NVIDIA Jetson AGX Orin development kit where the SoC module is combined with a carrier board. The system is powered by a USB-C connector, which is routed through PowerSensor3.

\begin{figure}
\centering
\includegraphics[width=1.0\columnwidth]{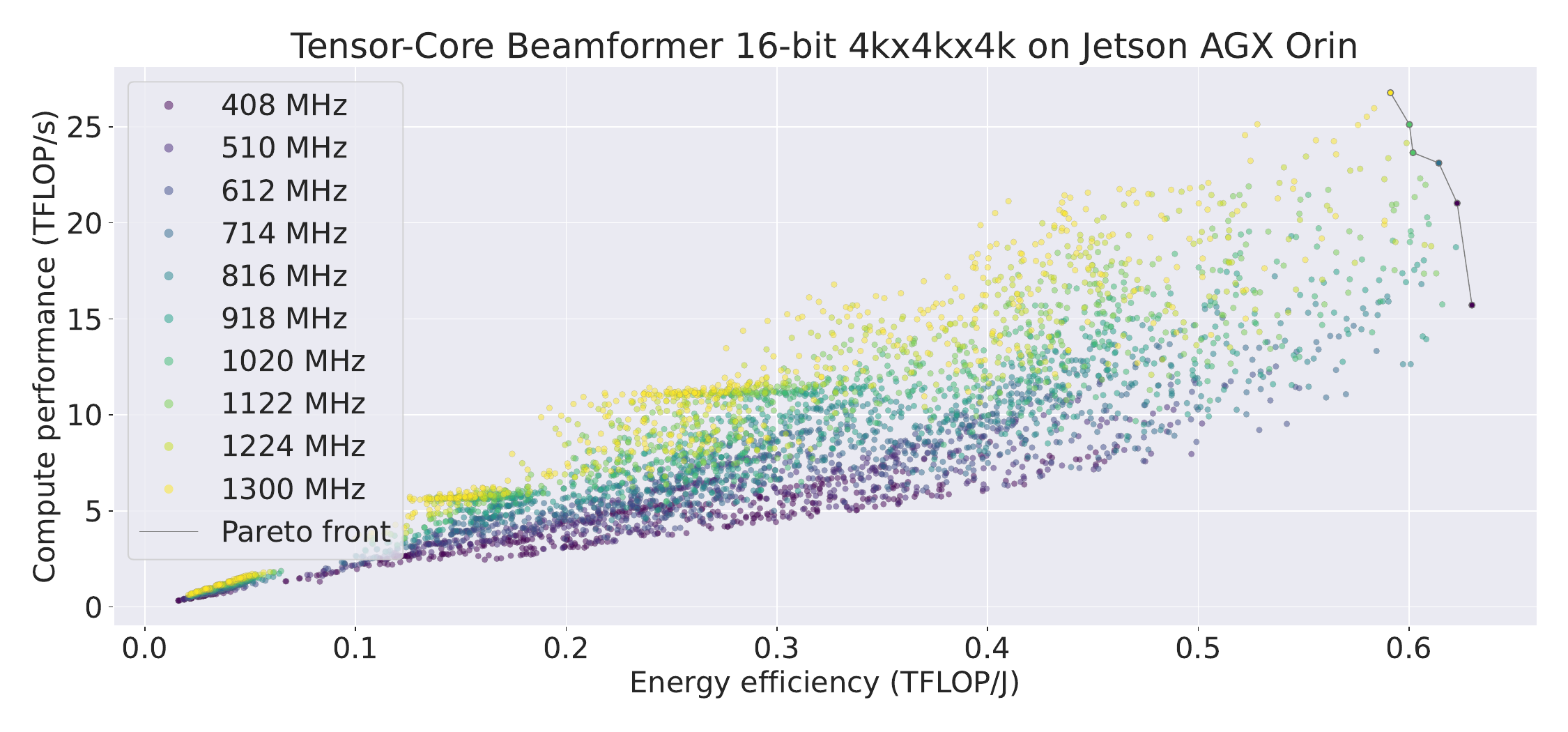}
\vspace{-0.7cm}
\caption{Tuning results for the Tensor-Core Beamformer on the NVIDIA Jetson AGX Orin.}\label{fig:beamformer-jetson}
\end{figure}

We repeat the same measurement as on the RTX 4000 Ada (\cref{fig:beamformer-ad4000}). The tuning results are shown in \cref{fig:beamformer-jetson}. The overall behavior is similar to the RTX 4000 Ada. PowerSensor3 provides several advantages over the built-in sensor of the Jetson: the time resolution of the built-in sensor is very limited ($\sim$0.1 second), and similar to the RTX 4000 Ada we can perform this experiment much faster with PowerSensor3. Additionally, the built-in sensor only measures the power consumption of the Jetson module, not including the carrier board that the module is inserted into. With PowerSensor3, we are able to measure the power consumption of the entire device. 

\subsection{SSDs}

\begin{figure}
    \centering
    \includegraphics[width=1.0\columnwidth]{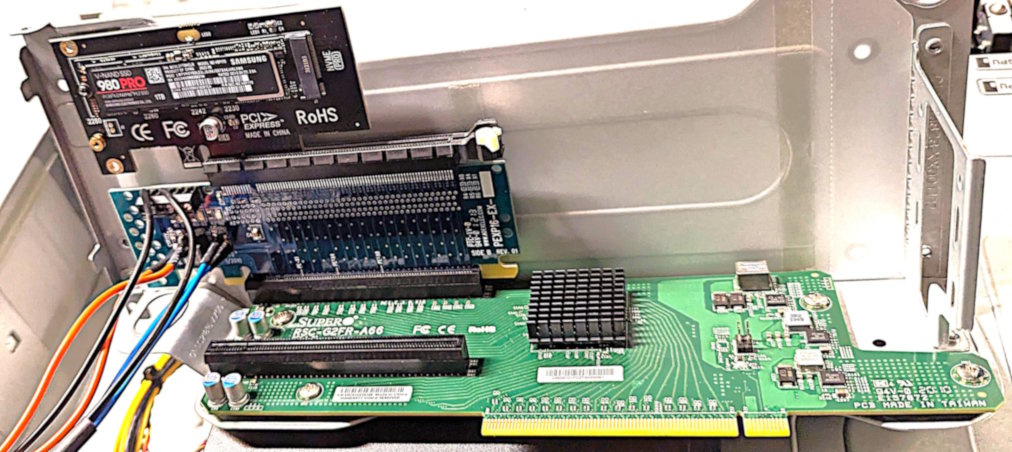} 
    \caption{The NVMe to PCIe adapter with the Samsung 980 PRO 1\,TB M.2 SSD~\cite{samsung-980-pro-spec}  in a modified PCIe (gen 3) riser card, providing measurement points for 3.3\,V and 12\,V.}\label{fig:SSD}
\end{figure}

Apart from GPUs, another major power consumer in data centers is data storage~\cite{rao2024understanding}, with power usage estimates ranging from 10\%~\cite{shehabi2016united,xie2024can} to as much as 25--30\%~\cite{caoanswering}.
In order to reduce the power utilization of storage, it is important to understand the power contribution of individual hardware components, such as individual SSDs.
However, storage devices do not report their power usage and rely on external sensors.
In this section, we demonstrate that the PowerSensor3 is an effective external sensor for modern storage devices.

Numerous investigations have been conducted to measure SSD power consumption,
%we can categorize these as:
categories related to this work are:
\begin{itemize}
    \item Individual flash chips and SATA drives~\cite{bjorling2010uflip, cho2015design, grupp2009characterizing, seo2008empirical}.
    \item Whole system energy of software on NVMe~\cite{2020-ull-ssds-energy, 2022-hotstorage-poll-energy, sundar2023energy, whitaker2023we}.
    \item Analysis of NVMe SSDs, using a custom external sensor~\cite{xie2024can}, which samples at 1\,kHz.
\end{itemize}
However, these studies either do not measure power of individual SSDs, lack standardized tooling, can not measure at the desired granularity, or do not apply to NVMe flash SSDs.
PowerSensor3 allows for a standardized approach for SSDs with a configurable granularity in sample frequency (sub-milliseconds to seconds). 

For evaluation we use a Samsung 980 PRO 1 TB M.2 SSD with the hardware setup visualized in \cref{fig:SSD}. The Supermicro system (SYS-2029GP-TR) used in this set-up does not easily provide access to PCIe slot power. Therefore, we use an additional PCIe 3.0 riser card, modified similar to the situation described for GPU measurements, providing measurement points for the PCIe 3.3\,V and 12\,V power channels.
We use the state-of-the-practice fio workload generator~\cite{2024-fio} with direct I/O and the \emph{io\_uring} engine with recommended performance optimizations~\cite{2023-cheops-storagestackcharacterization}.
As demonstrative workloads, we use random reads at various request sizes and use a long-running random write workload.

\begin{figure*}
\centering
\begin{minipage}[b]{0.9\textwidth}
    \subfloat[Random reads, mean per request size.]{\hspace{-0.3cm}
        \includegraphics[width=0.5\linewidth]{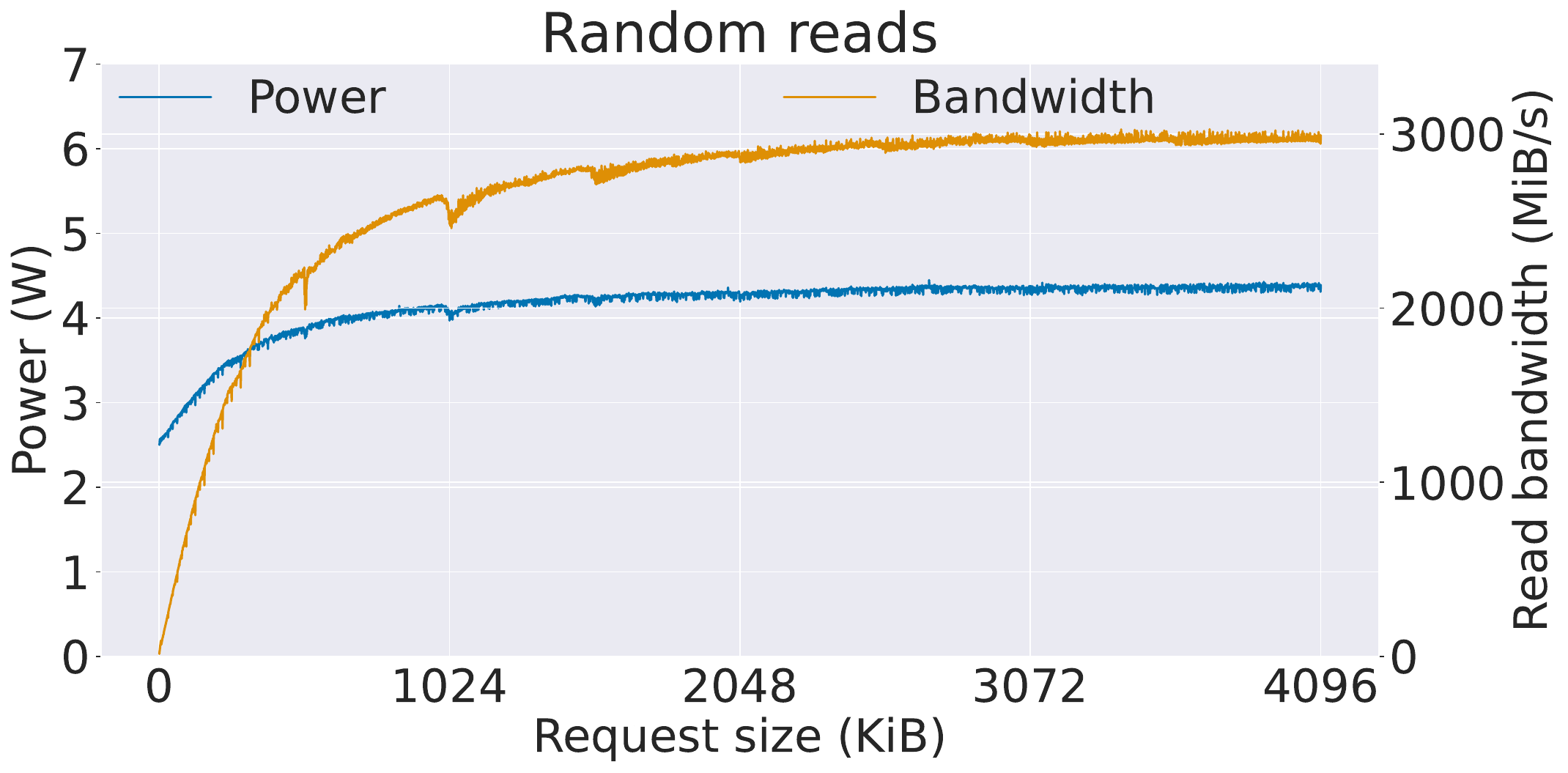}
        \label{fig:ssd-power-measurement-blocksize}}
    \subfloat[Random writes, mean over time.]{%\hspace{-0.2cm}
        \includegraphics[width=0.5\linewidth]
        {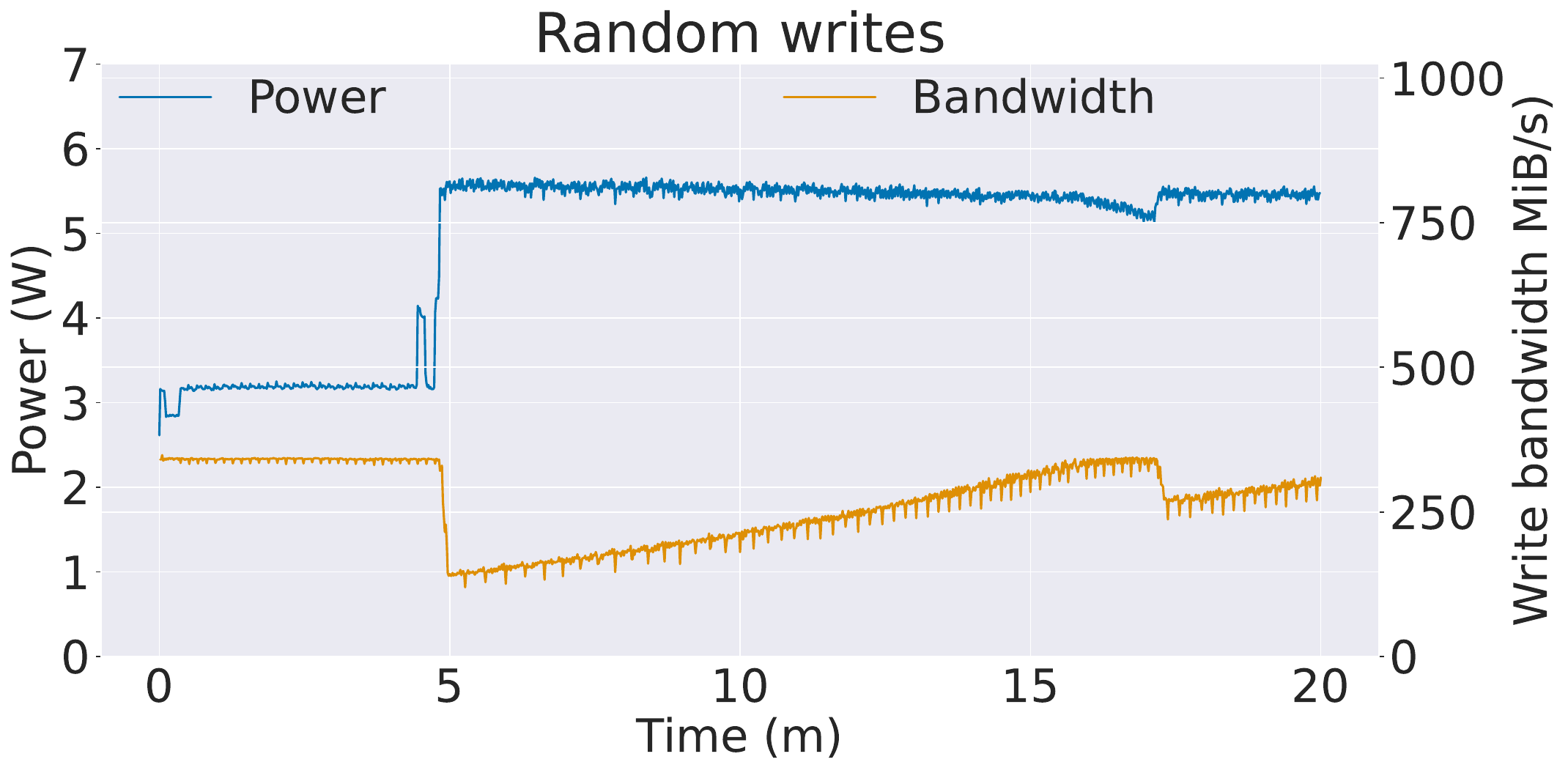}
        \label{fig:ssd-power-measurement-gc}}
        \caption{Power and bandwidth results for the Samsung 980 PRO benchmarking: (a) random reads; (b) random writes.}
        \label{fig:ssd-measurements}
\end{minipage}
\end{figure*}

First, we evaluate the impact of I/O request size for random reads on bandwidth and power.
It is well-known that larger requests typically lead to increased SSD bandwidth and power~\cite{xie2024can, cho2015design} as more work can be done in parallel.
To reproduce these observations, we run 10 second long random read workloads at request sizes ranging from 1--4096\,KiB ($\Delta$1\,KiB).
In \cref{fig:ssd-power-measurement-blocksize}, we plot the read request size on the x-axis, and the average power usage and bandwidth on the y-axes.
We confirm that power and bandwidth both increase with the request size (expected) until the device is saturated. 

Second, we evaluate the bandwidth and power for a longer-running ($>$20\,minute) random write workload.
Flash SSDs (with a block interface) are known to suffer from performance variability when consistently writing randomly, which is largely due to an SSD-internal process known as garbage collection (GC).
GC issues reads and writes that interfere with reads and writes issued by the host, which leads to performance variability.
Past studies have observed that this variability does not necessarily translate to similar trends in SSD power~\cite{seo2008empirical, cho2015design}.
Such discrepancies have implications for host-managed solutions that attempt to reduce SSD power or estimate SSD power usage, since bandwidth is not an accurate indicator of power.
We evaluate if these observations also hold for the evaluated SSD.
We first format the NVMe SSD, then precondition with 128\,KiB sequential writes, and lastly issue random 4\,KiB writes until the SSD is in steady-state.
\cref{fig:ssd-power-measurement-gc} shows the power and bandwidth (y-axes) over time (x-axis) for the random writes, using a granularity of one second for both power and bandwidth.
We observe that bandwidth is highly variable, but power increases to 5\,W at the first bandwidth descend, and remains relatively stable afterward.
We thus confirm that bandwidth is not indicative of power consumption. Therefore, to accurately evaluate SSD power for a given workload, we recommend using an external sensor.

To conclude, the PowerSensor3 allows us to reproduce prior SSD energy measurements, but with the added advantage that the sensor is standardized and can be installed within servers (deployment flexibility).
While we have evaluated our storage workloads at 1\,ms granularity, the PowerSensor3 is able to measure at sub-millisecond granularity (e.g., $>$1\,kHz) which will be evaluated in more detail in future work.

\section{Discussion}
\label{sec:discussion}
\textbf{Calibration and ease of use}:
The calibration and evaluation of the sensors, as described in this work, follow standard procedures for such devices. Calibration is required only once during production, ensuring long-term reliability and accuracy. The source code repository includes comprehensive documentation detailing the productions and calibration process, making it accessible for users to understand and implement, when they wish to produce their own hardware. Despite the open nature of the device and accompanying software, we acknowledge that not everyone may be able to manufacture the device independently. To address this, we have started an initiative to explore if we can provide fully assembled and calibrated devices, allowing for broader accessibility and ease of use of the PowerSensor3 technology \cite{CFPS}.

\textbf{Extendibility of PowerSensor3}:
The current design of PowerSensor3 allows to measure up to four different power supplies to a device, which can range from GPU cards and SoC boards to custom boards with ASICs. The provided software is compatible with any host system running Linux, offering flexibility and adaptability to various use cases. Both the hardware and software can be tailored to specific requirements such as different power ranges, sensor accuracy, connectors types and form factor. We encourage others to make their sensor boards available under an open hardware license and to open pull requests on the hardware and software repositories.

\section{Conclusion}
\label{sec:conclusions}
The application case studies presented in \cref{sec:application-case-studies} illustrate that PowerSensor3 offers an open, cost-efficient solution for fine-grained, high-frequency power measurements on a variety of peripheral devices. Thus, enabling a deeper understanding of system-level energy consumption and guiding effective optimization strategies.

For the NVIDIA RTX 4000 Ada GPU, PowerSensor3 reveals previously undetected behavior, outperforming built-in sensors. For the AMD W7700 GPU, it shows comparable time and amplitude accuracy to the built-in sensor, which specifications are not well documented. PowerSensor3 also reduced the Tensor-Core Beamformer application auto-tuning time by 3.25x compared to NVIDIA's internal sensor.

Additionally, PowerSensor3 works with SoC boards like the NVIDIA Jetson AGX Orin, which lacks total system power reporting. The power consumption of PowerSensor3 itself is minimal, measured in milliwatts, which is negligible compared to SoC boards. Alternatively, the PowerSensor3 can be powered separately, eliminating the need to draw power from the monitored system.

A case study with a PCIe SSD demonstrated that PowerSensor3 uncovers behavior not observable from bandwidth metrics alone, proving its utility for PCIe devices without built-in sensors.

PowerSensor3 allows developers and researchers to use energy as a metric for software optimization and evaluate the efficiency of new hardware platforms. PowerSensor3 can play a pivotal role in reducing the energy footprint of large-scale AI, HPC, and data center operations.

\ifdefined\DOUBLEBLIND

\else % if NOT double blind

\section*{Acknowledgment}

We would like to thank Quinten Twisk for his work on an early prototype of PowerSensor3.

\fi

\bibliographystyle{IEEEtran}
\bibliography{references}
\clearpage
% LaTeX template for Artifact Evaluation V20240722
%
% Prepared by Grigori Fursin with contributions from Bruce Childers,
%   Michael Heroux, Michela Taufer and other colleagues.
%
% See examples of this Artifact Appendix in
%  * ASPLOS'24 "PyTorch 2: Faster Machine Learning Through Dynamic Python Bytecode Transformation and Graph Compilation": 
%      https://dl.acm.org/doi/10.1145/3620665.3640366
%  * SC'17 paper: https://dl.acm.org/citation.cfm?id=3126948
%  * CGO'17 paper: https://www.cl.cam.ac.uk/~sa614/papers/Software-Prefetching-CGO2017.pdf
%  * ACM ReQuEST-ASPLOS'18 paper: https://dl.acm.org/citation.cfm?doid=3229762.3229763
%
% (C)opyright 2014-2024 cTuning.org
%
% CC BY 4.0 license
%

\appendix

\section{Artifact Evaluation Appendix}
Appendix containing Artifact description.

%%%%%%%%%%%%%%%%%%%%%%%%%%%%%%%%%%%%%%%%%%%%%%%%%%%%%%%%%%%%%%%%%%%%%
\subsection{Abstract}
In this work, we introduce the PowerSensor3, a novel tool comprising custom-developed hardware, firmware, and software components.

The hardware architecture of the PowerSensor3 includes a base board and a sensor board, both of which have been meticulously designed and released as open hardware under the CERN Open Hardware License (CERN-OHL-P v2) at \url{https://doi.org/10.5281/zenodo.15023417}~\cite{PS3-hw}. This open hardware approach ensures transparency, reproducibility, and the potential for community-driven enhancements.

Complementing the hardware, the firmware and host software for the PowerSensor3 have been developed and released as open-source software under the Apache License 2.0 at \url{https://doi.org/10.5281/zenodo.7941162}~\cite{PS3-sw}. This open-source software framework facilitates seamless integration with existing systems and promotes collaborative development.

To validate the performance and accuracy of the PowerSensor3, we conducted extensive evaluations using the Power Measurement Toolkit (PMT)~\cite{Corda2022} and KernelTuner~\cite{vanwerkhoven2019kernel}. These tools were employed in conjunction with a Tensor Core Beamformer application~\cite{oostrum2025tcbf} and storage benchmarks, enabling comprehensive analysis and benchmarking.

The results and findings from these evaluations are made available at \url{https://doi.org/10.5281/zenodo.15037450}~\cite{PS3-results} and 
\url{https://doi.org/10.5281/zenodo.15019310}~\cite{PS3-results-SSD}, both licensed with Apache 2.0. By providing access to these results, we aim to foster further research and development in the field of power measurement.

%%%%%%%%%%%%%%%%%%%%%%%%%%%%%%%%%%%%%%%%%%%%%%%%%%%%%%%%%%%%%%%%%%%%%
\subsection{Description}
\begin{figure}[h]
\centering
\includegraphics[width=\columnwidth]{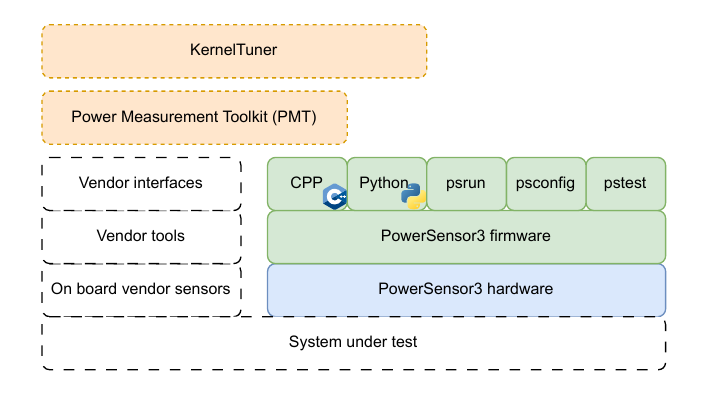}
\caption{Overview of components involved in the PowerSensor3 design and evaluation.}
\label{fig:PS3_stack}
\end{figure}

\cref{fig:PS3_stack} illustrates the organization of the various components in the PowerSensor3 hardware and software stack:

{\bf System Under Test (SUT):} The SUT, which can be a CPU, GPU, SoC, or other hardware components, often comes equipped with its own sensors, tools, and interfaces which can be compared to or combined with the PowerSensor3 measurements. These sensors, tools and interfaces are not part of this work, but are (when available) used in comparison to our PowerSensor3. 

\begin{figure}[h]
\centering
\includegraphics[width=0.5\textwidth]{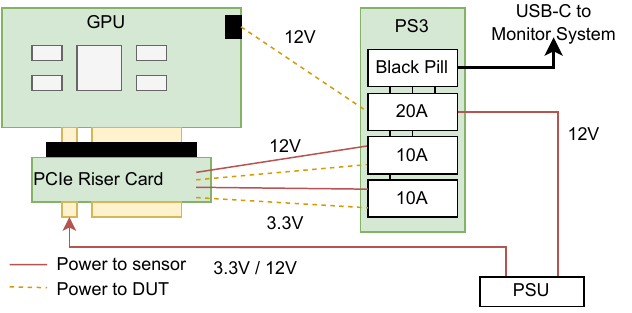}
\caption{Schematic of a PowerSensor3 measurement setup for PCIe devices.}
\label{fig:PS3_Schematic_Overview_ae}
\end{figure}

{\bf PowerSensor3 hardware:} The PowerSensor3 hardware \cite{PS3-hw}, shown in blue in \cref{fig:PS3_stack}, consists of the baseboard and sensor modules. The baseboard houses the STM32F411 microcontroller and supports up to four sensor modules, which can be customized to measure different power ranges and types of connectivity. \cref{fig:PS3_Schematic_Overview_ae} illustrates an example of the PowerSensor3 in operation. In this example, the PowerSensor3 is equipped with a PCIe sensor module that measures the power supplied to the PCIe card via the external power input as well as two sensor modules to measure the 3.3\,V and 12\,V PCIe slot power. A modified riser card, where the power connections for both 3.3\,V and 12\,V are interrupted and routed through two sensor modules, enables to measure the power consumption of the PCIe slot. The PowerSensor3 hardware design has been released as open hardware under the CERN Open Hardware License (CERN-OHL-P v2) at \url{https://doi.org/10.5281/zenodo.15023417} \cite{PS3-hw}.

{\bf PowerSensor3 firmware and software:} The PowerSensor3 firmware and software~\cite{PS3-sw}, depicted in green in \cref{fig:PS3_stack}, play a vital role in the system's functionality. The firmware is programmed onto the PowerSensor3 hardware, enabling it to process and transmit power measurement data. The host system interacts with the hardware through the PowerSensor3 software, which provides a user-friendly interface for configuring, monitoring, and analyzing power consumption data. For easy use, the executable \emph{psrun} can be used with a command line interface to report the power utilization of an existing application. For measurements with very high time resolution, Python and CPP interfaces are offered for integration with user applications. The PowerSensor3 firmware and software have been released as open-source software under the Apache License 2.0 at \url{https://doi.org/10.5281/zenodo.7941162}~\cite{PS3-sw}.

{\bf Power Measurement Toolkit (PMT) and KernelTuner:} PMT~\cite{Corda2022} and KernelTuner~\cite{vanwerkhoven2019kernel}, shown in orange in \cref{fig:PS3_stack}, are versatile tools that can be used with both vendor-specific sensors and the PowerSensor3. PMT is designed for comprehensive power measurement and analysis across various devices, while KernelTuner facilitates the optimization of GPU kernel performance across a broad range of parameters. These tools enhance the capability to evaluate and fine-tune the power consumption and efficiency of the SUT. PMT and KernelTuner are not part of this work, but are used in evaluation of our PowerSensor3. We have contributed the PowerSensor3 specific extensions of PMT and Kernel Tuner back to these projects.

For evaluation of PowerSensor3 with Kernel Tuner we carefully selected representative kernels that align with real-world high-performance GPU workloads where power efficiency is a critical concern. While vendor-provided reference implementations (e.g., CUTLASS and cuBLAS) may serve as performance baselines, our goal was to analyze power behavior in a use case from our application domains. The speedup achieved in tuning these specific kernels with Kernel Tuner are similar with the PowerSensor3 and the vendor tooling, however the tuning itself required 3.25$\times$ less time with PowerSensor3.

{\bf Evaluation results and SSD dataset:} The results presented in this work~\cite{PS3-results} and the supplementary SSD dataset~\cite{PS3-results-SSD} provide valuable insights into the performance and accuracy of the PowerSensor3. These datasets are made available for evaluation purposes and include detailed examples on how to effectively utilize the PowerSensor3 for various applications. By sharing these results, we aim to support further research and development in power measurement technologies. The results and findings from these evaluations are made available at \url{https://doi.org/10.5281/zenodo.15037450}~\cite{PS3-results} and 
\url{https://doi.org/10.5281/zenodo.15019310}~\cite{PS3-results-SSD}.

\begin{figure}[h!]
\centering
\includegraphics[width=0.45\textwidth]{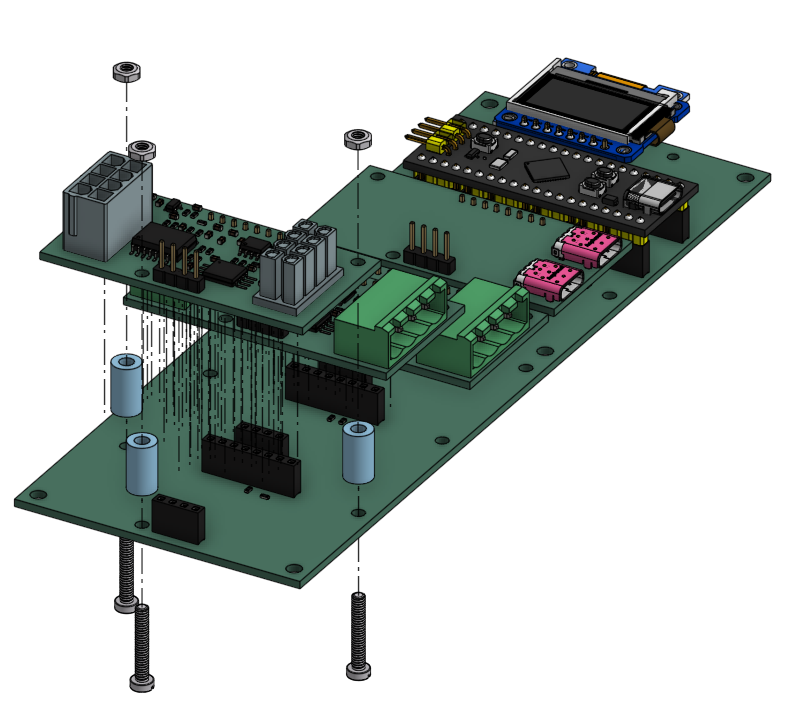}
\caption{Assembly instructions for the PowerSensor3 baseboard and sensor modules.}
\label{fig:PS3-assembly}
\end{figure}

\begin{figure}[h!]
\centering
\includegraphics[width=0.45\textwidth]{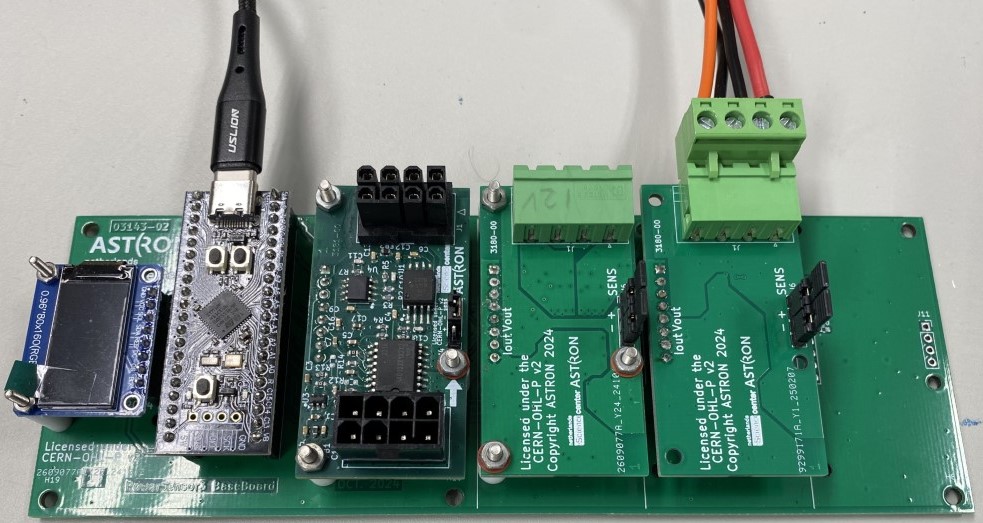}
\caption{Assembled PowerSensor3 baseboard with three sensor modules populated.}
\label{fig:PS3-assembled}
\end{figure}

{\bf Documentation:} The installation and use of the PowerSensor3 hardware, firmware and software is documented at: \url{https://powersensor3.readthedocs.io/en/latest/} \cite{PS3-doc} and described in readme files in the individual repositories. \cref{fig:PS3-assembly} shows an example of assembly instructions as found in the hardware repository and \cref{fig:PS3-assembled} shows a fully assembled PowerSensor3 with three sensor modules populated.

{\bf Hardware dependencies:} The PowerSensor3 firmware and software depend on the PowerSensor3 hardware.

{\bf Software dependencies:} Software dependencies are described in the PowerSensor3 documentation \cite{PS3-doc} and are managed through git submodules and cmake files in the repository, the hardware design works with KiCAD and the software has been designed for Linux.

{\bf How to contribute:} We encourage others to contribute to the development of PowerSensor3. Possible forms of contributions include: integration of the PowerSensor3 library with your software, pull requests for extensions to the hardware, firmware, software and documentation and design of your own sensor boards, made available under an open hardware license.

\end{document}